\RequirePackage{lineno}
\documentclass[aps,prd,twocolumn,superscriptaddress,showpacs]{revtex4}
\usepackage{epsfig}
\usepackage{graphicx}
\usepackage{dcolumn}
\usepackage{amsmath}
\usepackage{subfigure}
\usepackage{xspace}
\usepackage{color}
\usepackage{amssymb}
\usepackage{epsfig}
\usepackage{footnote}
\usepackage{longtable}
\usepackage{fancyhdr}
\usepackage{subfigure}
\usepackage{xspace}
\usepackage{multirow}
\usepackage{comment}
\usepackage{ifthen}

\usepackage{lineno}

\newboolean{PRD}
\newboolean{Thesis}
\setboolean{PRD}{true}
\setboolean{Thesis}{false}

\newcommand{\Section}[1]{Section~\ref{#1}}

\def \epage {\enlargethispage*{2.0in}}
\def \epage2 {\enlargethispage*{2.0in}}
\def \epage5 {\enlargethispage*{5.0in}}
\def \bc {\begin{center}}
\def \ec {\end{center}}

\def \Del \nabla
\def \del \partial
\def \hbar {\not h}
\def \ppm $\pm $
\def \D0 {D\O}

\def \Et {{\rm E}_{\rm T}}

\def\degrees{^\circ}

\def\GeVc2{GeV/{c^2}}

\def\nb-1{nb^{-1}}
\def\pb-1{pb^{-1}}
\def\fb-1{fb^{-1}}

\def\Z{${\em Z }$}

\def\Z0{{ Z^0}}

\def\95cl{95 \%~C.L.}
\def\95CL{95 \%~C.L.}
\def\r#1 {$^{#1}$}

\def\M4j{M_{4j}}
\def\M12{M_{12}}
\def\M34{M_{34}}

\newcommand{\met}{\mbox{${\not\!\!E_T}$}}

\newcommand{\degs}{\mbox{$^{\circ}$}}

\newcommand{\metvec}{{\not\!\! \vec{E}_T}}

\begin{document}


\title{Search for anomalous production of multiple leptons in association with $W$ and $Z$ bosons at CDF}
\affiliation{Institute of Physics, Academia Sinica, Taipei, Taiwan 11529, Republic of China}
\affiliation{Argonne National Laboratory, Argonne, Illinois 60439, USA}
\affiliation{University of Athens, 157 71 Athens, Greece}
\affiliation{Institut de Fisica d'Altes Energies, ICREA, Universitat Autonoma de Barcelona, E-08193, Bellaterra (Barcelona), Spain}
\affiliation{Baylor University, Waco, Texas 76798, USA}
\affiliation{Istituto Nazionale di Fisica Nucleare Bologna, $^{dd}$University of Bologna, I-40127 Bologna, Italy}
\affiliation{University of California, Davis, Davis, California 95616, USA}
\affiliation{University of California, Los Angeles, Los Angeles, California 90024, USA}
\affiliation{Instituto de Fisica de Cantabria, CSIC-University of Cantabria, 39005 Santander, Spain}
\affiliation{Carnegie Mellon University, Pittsburgh, Pennsylvania 15213, USA}
\affiliation{Enrico Fermi Institute, University of Chicago, Chicago, Illinois 60637, USA}
\affiliation{Comenius University, 842 48 Bratislava, Slovakia; Institute of Experimental Physics, 040 01 Kosice, Slovakia}
\affiliation{Joint Institute for Nuclear Research, RU-141980 Dubna, Russia}
\affiliation{Duke University, Durham, North Carolina 27708, USA}
\affiliation{Fermi National Accelerator Laboratory, Batavia, Illinois 60510, USA}
\affiliation{University of Florida, Gainesville, Florida 32611, USA}
\affiliation{Laboratori Nazionali di Frascati, Istituto Nazionale di Fisica Nucleare, I-00044 Frascati, Italy}
\affiliation{University of Geneva, CH-1211 Geneva 4, Switzerland}
\affiliation{Glasgow University, Glasgow G12 8QQ, United Kingdom}
\affiliation{Harvard University, Cambridge, Massachusetts 02138, USA}
\affiliation{Division of High Energy Physics, Department of Physics, University of Helsinki and Helsinki Institute of Physics, FIN-00014, Helsinki, Finland}
\affiliation{University of Illinois, Urbana, Illinois 61801, USA}
\affiliation{The Johns Hopkins University, Baltimore, Maryland 21218, USA}
\affiliation{Institut f\"{u}r Experimentelle Kernphysik, Karlsruhe Institute of Technology, D-76131 Karlsruhe, Germany}
\affiliation{Center for High Energy Physics: Kyungpook National University, Daegu 702-701, Korea; Seoul National University, Seoul 151-742, Korea; Sungkyunkwan University, Suwon 440-746, Korea; Korea Institute of Science and Technology Information, Daejeon 305-806, Korea; Chonnam National University, Gwangju 500-757, Korea; Chonbuk National University, Jeonju 561-756, Korea}
\affiliation{Ernest Orlando Lawrence Berkeley National Laboratory, Berkeley, California 94720, USA}
\affiliation{University of Liverpool, Liverpool L69 7ZE, United Kingdom}
\affiliation{University College London, London WC1E 6BT, United Kingdom}
\affiliation{Centro de Investigaciones Energeticas Medioambientales y Tecnologicas, E-28040 Madrid, Spain}
\affiliation{Massachusetts Institute of Technology, Cambridge, Massachusetts 02139, USA}
\affiliation{Institute of Particle Physics: McGill University, Montr\'{e}al, Qu\'{e}bec, Canada H3A~2T8; Simon Fraser University, Burnaby, British Columbia, Canada V5A~1S6; University of Toronto, Toronto, Ontario, Canada M5S~1A7; and TRIUMF, Vancouver, British Columbia, Canada V6T~2A3}
\affiliation{University of Michigan, Ann Arbor, Michigan 48109, USA}
\affiliation{Michigan State University, East Lansing, Michigan 48824, USA}
\affiliation{Institution for Theoretical and Experimental Physics, ITEP, Moscow 117259, Russia}
\affiliation{University of New Mexico, Albuquerque, New Mexico 87131, USA}
\affiliation{The Ohio State University, Columbus, Ohio 43210, USA}
\affiliation{Okayama University, Okayama 700-8530, Japan}
\affiliation{Osaka City University, Osaka 588, Japan}
\affiliation{University of Oxford, Oxford OX1 3RH, United Kingdom}
\affiliation{Istituto Nazionale di Fisica Nucleare, Sezione di Padova-Trento, $^{ee}$University of Padova, I-35131 Padova, Italy}
\affiliation{University of Pennsylvania, Philadelphia, Pennsylvania 19104, USA}
\affiliation{Istituto Nazionale di Fisica Nucleare Pisa, $^{ff}$University of Pisa, $^{gg}$University of Siena and $^{hh}$Scuola Normale Superiore, I-56127 Pisa, Italy}
\affiliation{University of Pittsburgh, Pittsburgh, Pennsylvania 15260, USA}
\affiliation{Purdue University, West Lafayette, Indiana 47907, USA}
\affiliation{University of Rochester, Rochester, New York 14627, USA}
\affiliation{The Rockefeller University, New York, New York 10065, USA}
\affiliation{Istituto Nazionale di Fisica Nucleare, Sezione di Roma 1, $^{ii}$Sapienza Universit\`{a} di Roma, I-00185 Roma, Italy}
\affiliation{Rutgers University, Piscataway, New Jersey 08855, USA}
\affiliation{Texas A\&M University, College Station, Texas 77843, USA}
\affiliation{Istituto Nazionale di Fisica Nucleare Trieste/Udine, I-34100 Trieste, $^{jj}$University of Udine, I-33100 Udine, Italy}
\affiliation{University of Tsukuba, Tsukuba, Ibaraki 305, Japan}
\affiliation{Tufts University, Medford, Massachusetts 02155, USA}
\affiliation{University of Virginia, Charlottesville, Virginia 22906, USA}
\affiliation{Waseda University, Tokyo 169, Japan}
\affiliation{Wayne State University, Detroit, Michigan 48201, USA}
\affiliation{University of Wisconsin, Madison, Wisconsin 53706, USA}
\affiliation{Yale University, New Haven, Connecticut 06520, USA}

\author{T.~Aaltonen}
\affiliation{Division of High Energy Physics, Department of Physics, University of Helsinki and Helsinki Institute of Physics, FIN-00014, Helsinki, Finland}
\author{B.~\'{A}lvarez~Gonz\'{a}lez$^y$}
\affiliation{Instituto de Fisica de Cantabria, CSIC-University of Cantabria, 39005 Santander, Spain}
\author{S.~Amerio}
\affiliation{Istituto Nazionale di Fisica Nucleare, Sezione di Padova-Trento, $^{ee}$University of Padova, I-35131 Padova, Italy}
\author{D.~Amidei}
\affiliation{University of Michigan, Ann Arbor, Michigan 48109, USA}
\author{A.~Anastassov$^w$}
\affiliation{Fermi National Accelerator Laboratory, Batavia, Illinois 60510, USA}
\author{A.~Annovi}
\affiliation{Laboratori Nazionali di Frascati, Istituto Nazionale di Fisica Nucleare, I-00044 Frascati, Italy}
\author{J.~Antos}
\affiliation{Comenius University, 842 48 Bratislava, Slovakia; Institute of Experimental Physics, 040 01 Kosice, Slovakia}
\author{G.~Apollinari}
\affiliation{Fermi National Accelerator Laboratory, Batavia, Illinois 60510, USA}
\author{J.A.~Appel}
\affiliation{Fermi National Accelerator Laboratory, Batavia, Illinois 60510, USA}
\author{T.~Arisawa}
\affiliation{Waseda University, Tokyo 169, Japan}
% ...

\collaboration{CDF Collaboration\footnote{With visitors from
$^a$Istituto Nazionale di Fisica Nucleare, Sezione di Cagliari, 09042 Monserrato (Cagliari), Italy,
$^b$University of CA Irvine, Irvine, CA 92697, USA,
$^c$University of CA Santa Barbara, Santa Barbara, CA 93106, USA,
$^d$University of CA Santa Cruz, Santa Cruz, CA 95064, USA,
$^e$Institute of Physics, Academy of Sciences of the Czech Republic, Czech Republic,
$^f$CERN, CH-1211 Geneva, Switzerland,
$^g$Cornell University, Ithaca, NY 14853, USA,
$^h$University of Cyprus, Nicosia CY-1678, Cyprus,
$^i$Office of Science, U.S. Department of Energy, Washington, DC 20585, USA,
$^j$University College Dublin, Dublin 4, Ireland,
$^k$ETH, 8092 Zurich, Switzerland,
$^l$University of Fukui, Fukui City, Fukui Prefecture, Japan 910-0017,
$^m$Universidad Iberoamericana, Mexico D.F., Mexico,
$^n$University of Iowa, Iowa City, IA 52242, USA,
$^o$Kinki University, Higashi-Osaka City, Japan 577-8502,
$^p$Kansas State University, Manhattan, KS 66506, USA,
$^q$University of Manchester, Manchester M13 9PL, United Kingdom,
$^r$Queen Mary, University of London, London, E1 4NS, United Kingdom,
$^s$University of Melbourne, Victoria 3010, Australia,
$^t$Muons, Inc., Batavia, IL 60510, USA,
$^u$Nagasaki Institute of Applied Science, Nagasaki, Japan,
$^v$National Research Nuclear University, Moscow, Russia,
$^w$Northwestern University, Evanston, IL 60208, USA,
$^x$University of Notre Dame, Notre Dame, IN 46556, USA,
$^y$Universidad de Oviedo, E-33007 Oviedo, Spain,
$^z$CNRS-IN2P3, Paris, F-75252 France,
$^{aa}$Texas Tech University, Lubbock, TX 79609, USA,
$^{bb}$Universidad Tecnica Federico Santa Maria, 110v Valparaiso, Chile,
$^{cc}$Yarmouk University, Irbid 211-63, Jordan,
}}
\noaffiliation

\date{\today}

\ifthenelse{\boolean{Thesis}}
{
\abstract
}
{
  \begin{abstract}
}

This paper presents a search for anomalous production of multiple
low-energy leptons in association with a $W$ or $Z$ boson using events
collected at the CDF experiment corresponding to $5.1$ fb$^{-1}$ of
integrated luminosity. This search 
is sensitive to a wide range of topologies
with low-momentum leptons, including those with the leptons near one
another.  The observed rates of production of additional electrons and
muons are compared with the standard model predictions. No indications
of phenomena beyond the standard model are found. A 95\% confidence
level limit is presented on the production cross section for a
benchmark model of supersymmetric hidden-valley Higgs
production.  Particle identification efficiencies are also provided 
to enable the calculation of limits on additional models.

\ifthenelse{\boolean{PRD}}
{
  \end{abstract}
}
{}


\maketitle

\section{Introduction}
\label{sec:introduction}

\ifthenelse{\boolean{Thesis}}
{The standard model (SM) of particle physics has been very successful in predicting observations made at high-energy particle colliders.  However, it is known to be incomplete.  There must be new physics beyond the reach of the current generation of experiments, and there are a plethora of theories predicting what that new physics might be.

Testing all of these models individually would be impossible, due to the number of models and the number of free parameters in each model.  Therefore, ``signature-based'' searches are performed: a signature is chosen that is common to many new models, but has a low rate of background production in the SM.  A search for this signature can both act as a precision test of the SM (by looking for disagreement with the low predicted background) and check for hints of many different possible theories of new physics.
}
{}

\begin{figure}[htb]
    \begin{center}
       \includegraphics[width=0.45\textwidth]{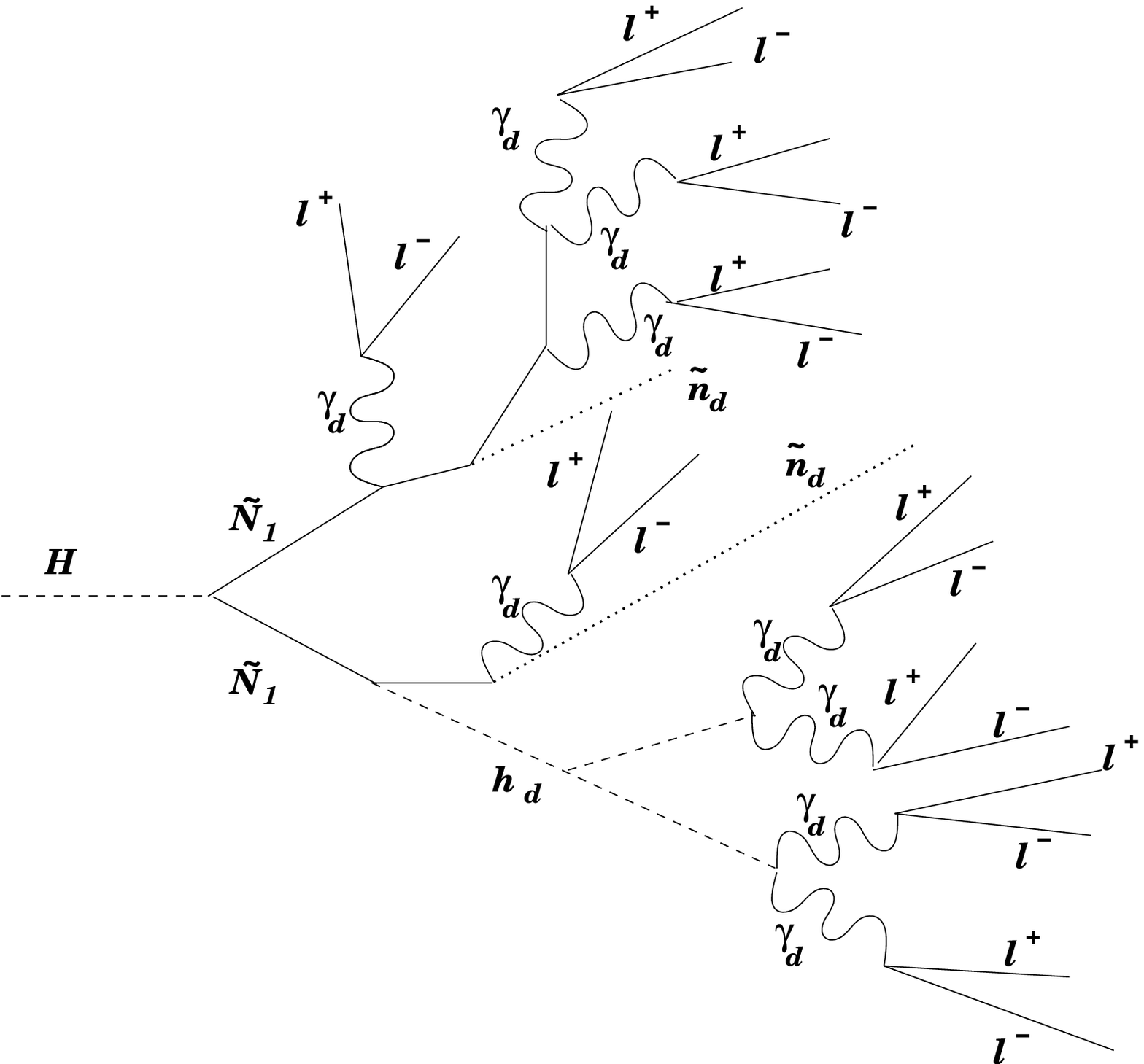}
        \caption{\label{fig:HHFeynman} An example of multiple low-$p_T$, non-isolated lepton production.  A Higgs decays to a pair of lightest supersymmetric neutralinos ($\tilde N_1$) which then cascade through a dark sector to a lightest dark sector particle ($\tilde n_d$) and a number of dark photons ($\gamma_d$).  The dark photons then decay back into the SM in the form of leptons ($l^\pm$).  This model is adapted from Ref.~\cite{HiggsToLeptonJets}.  Note that this diagram shows only the decay of the Higgs, while this analysis as a whole would be sensitive to the associated production of a Higgs with a $W$ or $Z$ boson.}
     \end{center}
\end{figure}

The signature of multiple leptons is common in many models of physics beyond the \ifthenelse{\boolean{PRD}}{standard model (SM)}{SM} with light mass scales and couplings to the electroweak sector, such as the Next-to-Minimal Supersymmetric Model~\cite{NMSSM}, little Higgs models~\cite{littleHiggs}, and  R-parity violating MSSM models~\cite{RparityMSSM}. Some of these
new physics scenarios propose explanations for the nature of dark matter~\cite{dmAnnihilation} as well as the existence of other, yet-undiscovered particles in long decay chains.  In addition to predicting large numbers of leptons, these models also often predict that clusters of leptons are produced spatially close to each other. These clusters are often referred to in the literature as ``lepton jets''~\cite{hiddenSectorDecays}.
Due to the unique characteristics of these models, they could have evaded previous searches for an excess of leptons, such as diboson searches~\cite{PrevDB} and SUSY-inspired multi-lepton searches~\cite{PrevSUSY}.  The high multiplicity of leptons can lead to low lepton momenta, well below the usual cutoff of 10-20 GeV.  Additionally, collimated lepton jets will fail the standard requirement that leptons be isolated in the detector.
\ifthenelse{\boolean{Thesis}}
{

  One of the recent
promising proposals involves the phenomenology of light supersymmetric hidden sectors~\cite{hiddensector} where the lightest
visible superpartner, the equivalent of the LSP in the MSSM, is allowed to cascade into a hidden sector.  The existence of such
sectors has been further motivated by recent observed astrophysical anomalies~\cite{astroAnomaly} which may be signatures of dark
matter annihilation~\cite{dmAnnihilation} or decays into a light hidden sector~\cite{hiddenSectorDecays}.

}
{}
As an example, Figure~\ref{fig:HHFeynman} shows a typical decay chain in a model in which the Higgs decays to a light hidden sector resulting in events with a high multiplicity of leptons~\cite{HiggsToLeptonJets}.

This \ifthenelse{\boolean{PRD}}{paper}{thesis} presents a signature-based search for anomalous production of multiple electrons and/or muons in association with $W$ and $Z$ bosons.  Previous searches for lepton jets at the Tevatron~\cite{PrevD0} and at the LHC~\cite{PrevCms} have focused on searching for clusters of leptons with specific requirements on the size of the clusters.  These searches have resulted in no evidence for lepton jets.  We have performed a more general search, sensitive to a wide range of scenarios that predict multiple electrons and muons.  Note that hadronic decays of tau leptons are not included in this search due to the additional difficulty in identifying them in non-isolated topologies.

The data used here correspond to $5.1$ fb$^{-1}$ of integrated luminosity at a center-of-mass energy of $\sqrt{s}=1.96$ TeV collected using the CDF detector at Fermilab between December 2004 and January 2010.  Within the events containing leptonically decaying $W$ and $Z$ bosons, we search for additional `soft' leptons with no isolation requirements and with momentum greater than $3$ GeV for muons and $2$ GeV for electrons~\cite{EtPt}.  

\section{Analysis strategy}
\label{sec:strategy}
The analysis strategy and the structure of this paper are as follows.
The baseline data sets for this analysis consist of leptonically decaying $W$ and $Z$ boson events selected with high transverse momentum~\cite{Coordinates} ($p_T$) leptons~\cite{InclusiveWZ}.  The kinematic distributions are used to validate the $W$ and $Z$ boson selections. 
The selection of these events is described in \Section{sec:evSelect}.
\ifthenelse{\boolean{Thesis}}{  A check of the ratio of cross-sections $\sigma(W)/\sigma(Z)$ is also used to validate the trigger efficiency and the lepton reconstruction efficiency.  This ratio $R$ of $W$ to $Z$ production is predicted to NNLO with a precision of a few percent~\cite{stirling_NNLO}, providing a very precise test of these efficiencies.}
{}

After the $W$ or $Z$ boson reconstruction, additional low-$p_T$ electrons and muons are identified in the events with no isolation requirements.
Purely data-driven techniques are used to develop the soft lepton identification algorithms.  The selection of soft leptons is more fully described in \Section{sec:SoftLep}.

The numbers of additional electrons and muons are counted in the inclusive $W$ and $Z$ data sets, where the SM predicts few events with multiple leptons.  The observed event count is compared to the SM expectations in bins of additional lepton multiplicity.
These results are described in \Section{sec:results}.

\section{The CDF II detector}
\label{sec:detector}
The CDF II detector is a cylindrically-symmetric
spectrometer designed to study $p \bar p$ collisions at the Fermilab
Tevatron. The detector has been extensively described in detail elsewhere in
the literature~\cite{CDF}.  \ifthenelse{\boolean{Thesis}}{An elevation view of the detector is shown in Figure~\ref{fig:detector}}{}  Here the
detector subsystems relevant for this analysis are described.

Tracking systems are used to measure the momenta of charged particles, to 
reconstruct primary and secondary vertices, and to trigger
 on and identify leptons with large transverse momentum. 
Silicon strip detectors (SVX)~\cite{SVX} and the central 
outer tracker (COT)~\cite{COT} are contained in a superconducting solenoid 
that generates a magnetic field of 1.4~T. 
The silicon strip system provides up to 8 measurements in the 
$r-\phi$ and $r-z$ views and helps to reconstruct tracks in the region $|\eta| <$ 2~\cite{Coordinates}. 
The COT is an open-cell drift chamber that makes up to 96 
measurements along the track of each charged particle in the
region $|\eta|<1$. Sense wires are arranged in 8 alternating axial
and $\pm 2\degs$ stereo super-layers. 
The resolution in $p_T$, $\sigma_{p_T}/p_T$, is 
$\approx 0.0015\;p_T ({\rm GeV})$ for tracks with only COT measurements, and 
$\approx 0.0007\;p_T ({\rm GeV})$ for tracks with both silicon and COT 
measurements.

Calorimeters are segmented with towers arranged in a projective geometry. 
Each tower consists of an electromagnetic and a hadronic
compartment~\cite{cem_resolution,cha,cal_upgrade}.  The central electromagnetic calorimeter (CEM) and central hadronic calorimeter (CHA) cover the central
region ($|\eta|< 1.1$), while the plug electromagnetic calorimeter (PEM) and plug hadronic
calorimeter (PHA) cover the `end plug' region ($1.1<|\eta|<3.6$).
In this analysis, a high-$E_T$ electron is
required to be identified in the central region, where the CEM has a
segmentation of  $15^{\degrees}$ in $\phi$ and $\approx 0.1$ in
$\eta$~\cite{CDF},
and an $E_T$ resolution of $\sigma(E_T)/E_T \approx
13.5\%/\sqrt{E_T (\rm{GeV})}\oplus 2\%$~\cite{cem_resolution}.  
Two additional systems in the central region with finer spatial
resolution are used for electron identification.  The
central strip system (CES) uses a multi-wire proportional chamber
to make profile measurements of electromagnetic showers at a depth
of 6 radiation lengths (approximately shower maximum)~\cite{cem_resolution}.  The central
preshower detector (CPR) is located just outside the solenoid coil on
the front face of the CEM.  In 2004 the CPR was upgraded 
from the Run I configuration of wire proportional chambers to a fast scintillator system~\cite{cal_upgrade}.  This analysis only uses data collected after the CPR upgrade.

Muons are identified using the central muon systems~\cite{muon_systems}: 
CMU and CMP for the pseudo-rapidity region of $|\eta|<0.6$, and
CMX for the pseudo-rapidity region of $0.6<|\eta|<1.0$. The CMU system uses 
four layers of planar drift chambers to detect muons with $p_T > 1.4$ GeV.
 The CMP system consists of an additional four layers
of planar drift chambers located behind 0.6 m of steel outside the
magnetic return yoke, and detects muons with $p_T > 2.2$ GeV. 
The CMX system detects muons with $p_T > 1.4$ GeV with four to
eight layers of drift chambers, depending on the direction of the muon.

The luminosity is measured using two sets of gas Cerenkov
counters~\cite{CLC}, located in the region $3.7<|\eta|<4.7$. The total
uncertainty on the luminosity is estimated to be 5.9\%, where
4.4\% comes from the acceptance and operation of the luminosity
monitor and 4.0\% from the calculation of the inelastic $p \bar p$
cross-section~\cite{LumiUncertainty}.

A three-level online event selection (trigger) system~\cite{trigger} selects events to be recorded
for further analysis. The first two trigger levels consist of
dedicated fast digital electronics analyzing a subset of the complete detector information.  The third level, applied to the full set of detector information from those
events passing the first two levels, consists of a farm of computers
that reconstruct the data and apply selection criteria 
consistent with the subsequent offline event processing.

\ifthenelse{\boolean{Thesis}}
{\begin{figure}
\includegraphics[angle=90, width=\textwidth]{plots/cdfelev.eps}
\caption{\label{fig:detector} An elevation view of the CDF II detector.  The detector is approximately symmetric around the collision point.}
\end{figure}}
{}

\section{$W$ and $Z$ boson sample selection}
\label{sec:evSelect}

Events for this analysis are selected with three different triggers~\cite{trigger}. Approximately half the events are selected with a trigger requiring
a high-$p_T$ central electron in the CEM ($E_T>18$ GeV, $|\eta|<1.0$). In addition, two muon triggers, one requiring hits in both the CMP and CMU and
the other requiring hits in the CMX, collect events with central muons ($p_T>18$ GeV, $|\eta|<1.0$). \ifthenelse{\boolean{Thesis}}{These datasets are described more fully in Appendix~\ref{sec:appendix_data}.}{}

Further selection criteria are imposed on triggered events offline.
\ifthenelse{\boolean{PRD}}{Electron (muon) candidates are required to have $E_T>20$ GeV ($p_T>20$ GeV). They must
fulfill several other identification criteria designed to select pure samples of high-$p_T$ electrons (muons)~\cite{InclusiveWZ}, including
an isolation requirement that the energy within a cone of $\Delta R=\sqrt{\Delta\phi^2 + \Delta\eta^2}<0.4$ around the lepton direction is less than 10\% of the $E_T$ ($p_T$) of 
the electron (muon).}
{The standard CDF selections~\cite{InclusiveWZ} are used to identify hard ($>20$ GeV) electrons and muons, with the additional requirement that hard muons have silicon hits.  Two categories of hard leptons are considered: `tight' leptons, which are described in Tables~\ref{tab:TCE_cuts} and \ref{tab:Muon_cuts}, are required as the trigger leg of the $W$ and $Z$.  A less stringent category of `loose' leptons, described in Tables~\ref{tab:Muon_cuts} and \ref{tab:LCE_cuts}, are required for the non-trigger leg of $Z$ reconstruction.}

In order to reduce the electron background from photon conversions, the electron(s) from the $W$ or $Z$ boson decay are required to pass a conversion filter.  \ifthenelse{\boolean{PRD}}{Electron candidates with an oppositely-charged partner track
consistent with having originated from a photon conversion are removed~\cite{MyThesis}.}{If there is an electron candidate that is oppositely-charged and has $\Delta\cot\theta < 0.04$ and $|\delta| < 0.2$, then
the electron is called a conversion. (See Fig.~\ref{fig:conv_vars} for an explanation and illustration of these variables.)}
However, the electron candidate is kept if its partner conversion track
also has another partner track, since the three tracks are assumed to
originate from an electron which radiates a photon which subsequently converts.

In order to reduce the background from mesons decaying to muons within the tracking chamber, the muon(s) from the $W$ or $Z$ boson decay must pass a decay-in-flight (DIF) removal algorithm. \ifthenelse{\boolean{PRD}}{The DIF algorithm requires the $\chi^2$ per degree of freedom of the fitted track to be less than 3 and
requires that the impact parameter of the track be less than 0.02 cm.}{A track left by a pion or a kaon that decayed to a muon within the detector will leave real hits in the muon detectors, but the track in the tracking chamber will have a noticeable `kink' in it where the decay occurred.  The DIF algorithm removes these tracks by requiring the $\chi^2$ per degree of freedom of the track fit to be less than 3 and the impact parameter of the track to be less than 0.02 cm.}  Additionally, for tracks
with $p_T > 300$ GeV, it requires $N_{\rm transitions} > 30$, where $N_{\rm transitions}$ is the number of times
the pattern of track hits crosses the fitted track~\cite{Sasha_DIF}. Muons consistent with cosmic rays are vetoed~\cite{CRVeto}.

\ifthenelse{\boolean{PRD}}
{
To select $W$ boson events we require $\met>25$ GeV and that the highest-energy lepton and the $\met$ have $m_{T}>20$ GeV~\cite{Coordinates}. In order to remove events where the $\met$ arises from a mismeasured lepton, the difference in $\phi$ between the
highest-energy lepton and the $\vec\met$ is required to be greater than 0.5 radians. The $Z$ boson selection requires two oppositely-charged, same-flavor leptons.  One of these
leptons is required to pass the above high-$p_T$ lepton identification selections while the other is required only to pass a less stringent `loose' selection. For muons, 
the loose selection allows for muons with $p_T\geq10$ GeV that have hits in either the CMP, CMU, or CMX systems. For electrons, the loose selection accepts
electrons with $E_T\geq12$ GeV and has relaxed identification requirements with respect to the centroid shape in the CES and $E/p$, the ratio of calorimeter energy to track momentum~\cite{HighPtID}.  Finally, the invariant mass of the lepton pair is required to be within the range of 76 GeV $\le m(l,l) \le$ 106 GeV, consistent with the mass of the $Z$ boson.  

The distributions of $m_T$ in $W$ boson events and the dilepton invariant mass in $Z$ boson events are shown in Figure~\ref{fig:WZ_valid}
for both electron- and muon-triggered events.  In total, 4,722,370 $W$ boson events and 342,291 $Z$ boson events are obtained from 5.1 fb$^{-1}$ of data.  Good agreement with predictions is observed across most of the distributions.  In the $W$ $m_T$ distributions, a disagreement occurs at low mass, where the distribution shifts from being QCD-dominated to electroweak-dominated, and is accounted for by the QCD normalization systematic uncertainty (as described in \Section{sec:Non_W}).  In the $Z$ selection, a similar mass disagreement is due to the fact that the Monte Carlo (MC) simulation does not include Drell-Yan events with a $Z/\gamma^*$ invariant mass below 8 GeV.  It is eliminated with the requirement that the dilepton mass be within the $Z$ peak.

\begin{figure}
\includegraphics[width=0.45\textwidth]{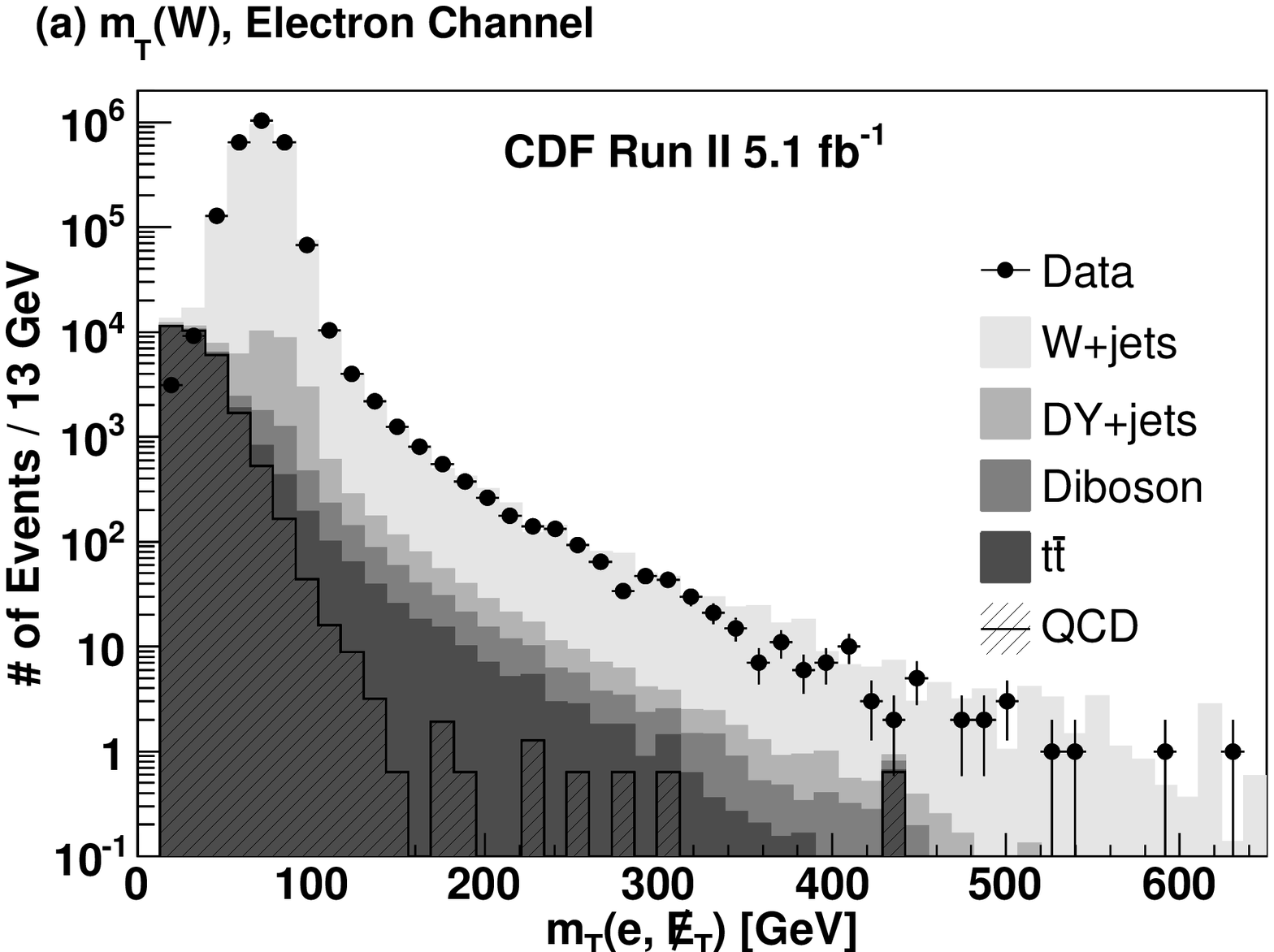}
\hfil
\includegraphics[width=0.45\textwidth]{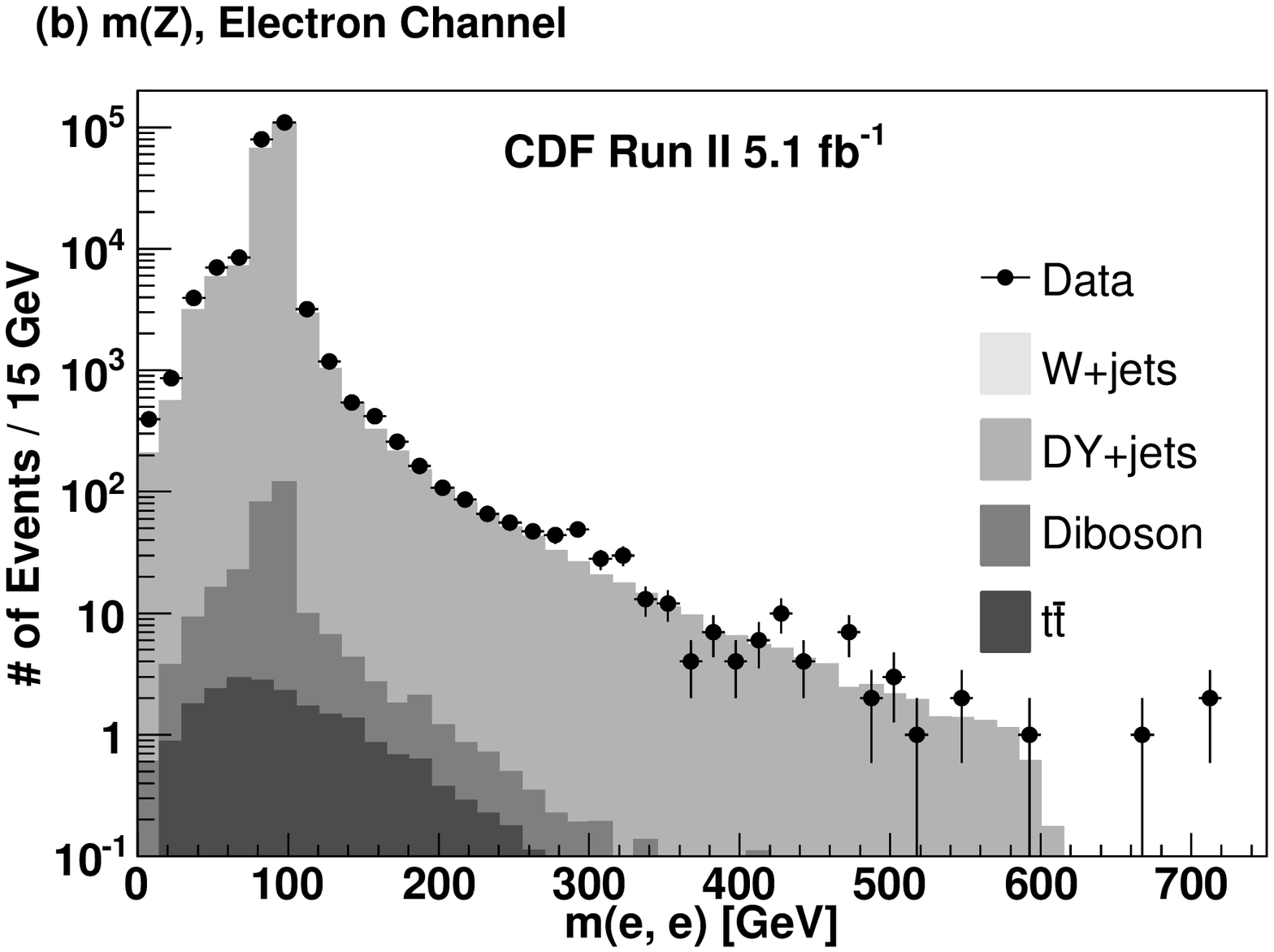}

\includegraphics[width=0.45\textwidth]{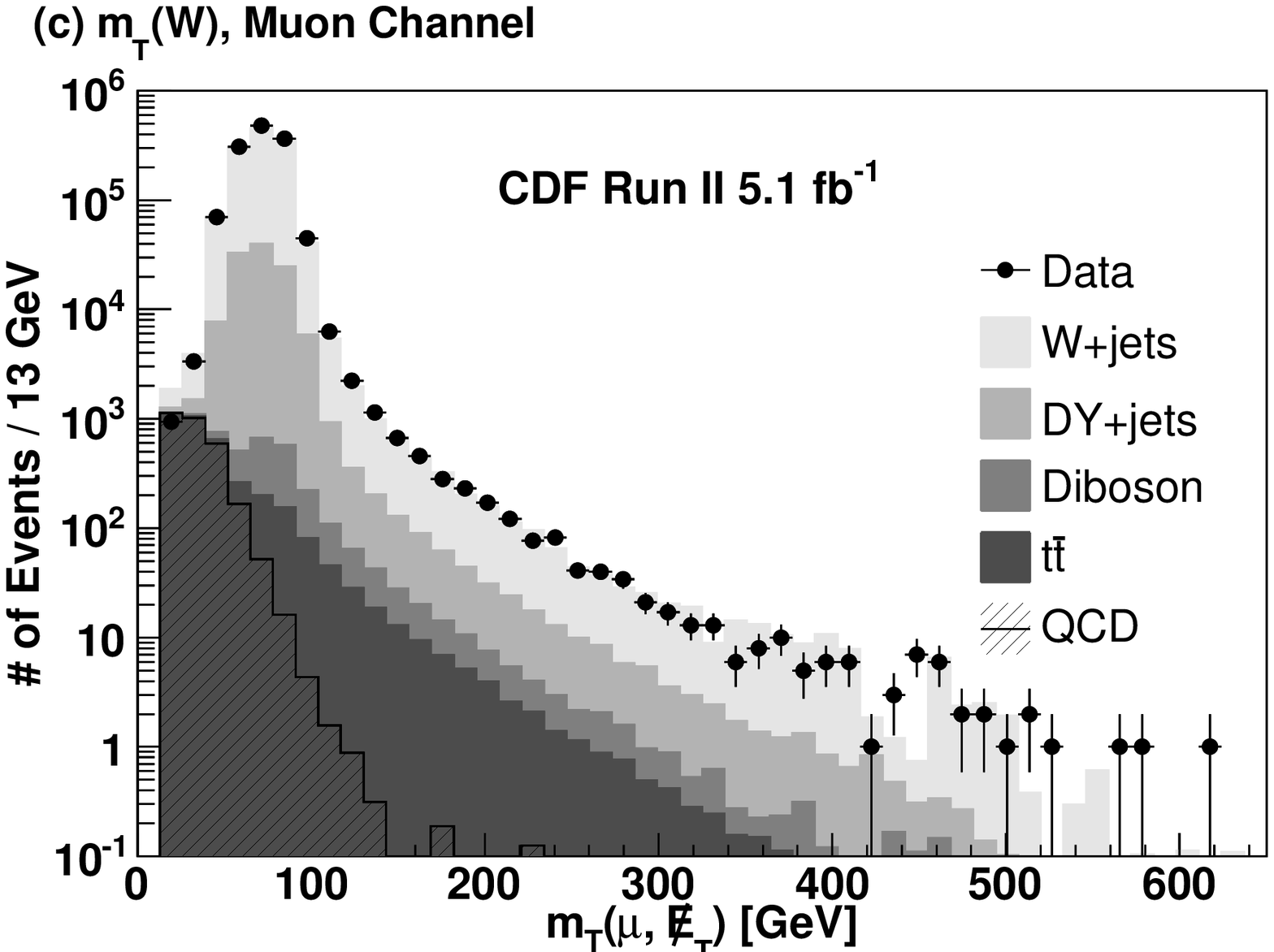}
\hfil
\includegraphics[width=0.45\textwidth]{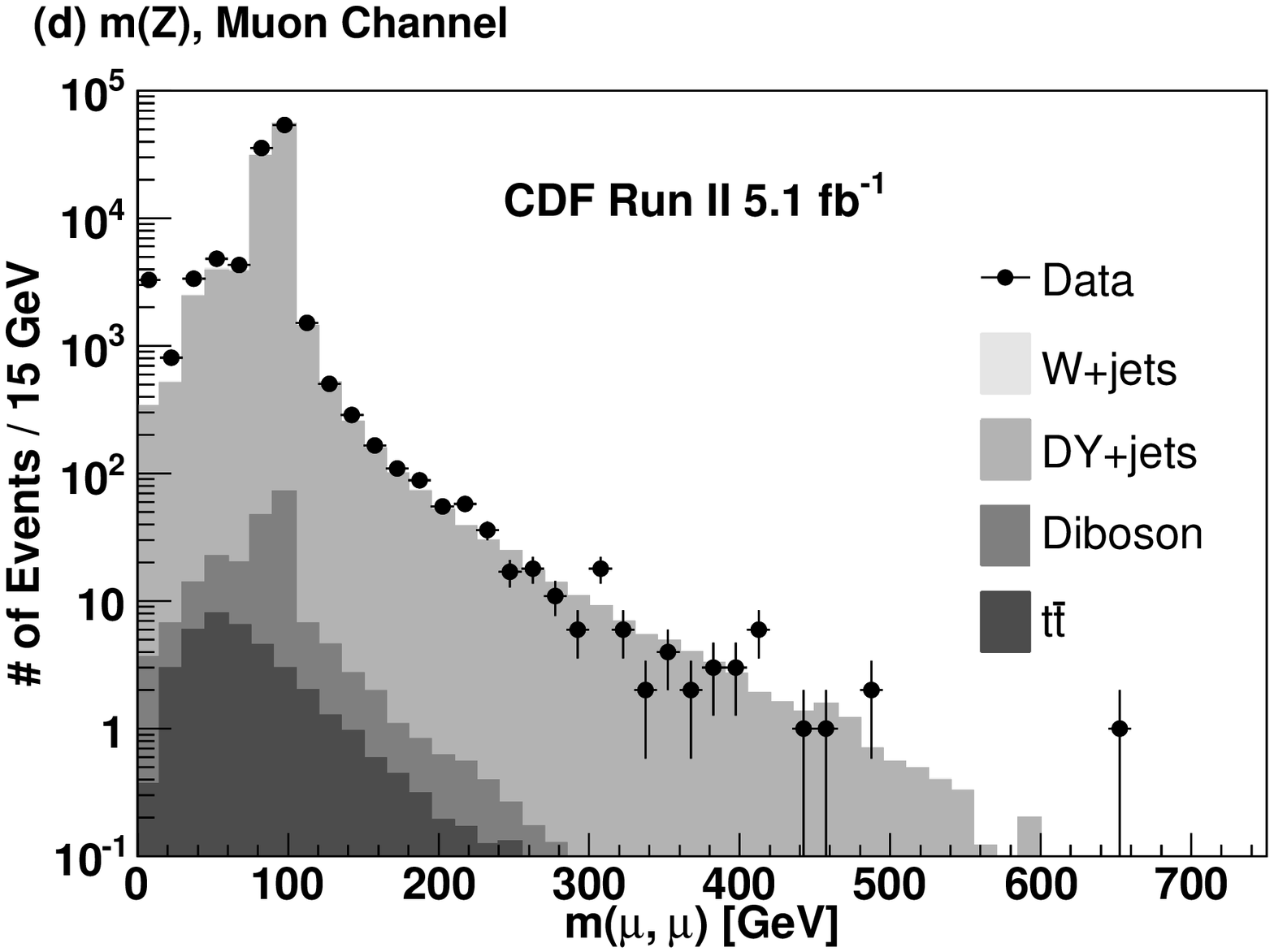}

\caption{\label{fig:WZ_valid} (a) The transverse mass ($m_T$) of the highest-$p_T$ lepton and the $\met$ in the electron-triggered $W$ boson sample. (b) The dilepton invariant mass in the electron-triggered $Z$ boson sample.  (c) The $m_T$ of the highest-$p_T$ lepton and the $\met$ in the muon-triggered $W$ boson sample. (d) The dilepton invariant mass in the muon-triggered $Z$ boson sample.
The estimation of the QCD contribution to these distributions is described in \Section{sec:Non_W}. The points represent the observed data and the filled histograms are the SM estimates.}
\end{figure}
}
{ 

\begin{table}
\centering
\begin{tabular}{|c|}
\hline
Tight Central Electron Selections \\
\hline
$\eta_\mathrm{Detector} \le 1.1$ \\
Track must be fiducial to CES \\
$E_T \ge 20$ GeV \\
$E_{HAD}/E_{EM} \le 0.055 + 0.00045 \times E$ \\
$\mathrm{Iso}(R=0.4)/E_T \le 0.1$ \\
$p_T \ge 10$ GeV \\
$\ge 3$ COT axial segments with $\ge 5$ hits \\
$\ge 2$ COT stereo segments with $\ge 5$ hits \\
$z_0^\mathrm{track} \le 60$ cm \\
$E/p \le 2$ unless $p_T \ge 50$ GeV \\
$\chi^2_{CES} \le 10$ \\
$-3.0$ cm $ \le Q \times \Delta X_{CES} \le 1.5$ cm \\
$|\Delta Z_{CES}|<3$ cm \\
$L_{shr} \le 0.2$ \\
Conversion Removal \\
\hline
\end{tabular}
\caption{\label{tab:TCE_cuts}Selections to identify tight central electrons}
\end{table}

\begin{table}
\centering
\begin{tabular}{|c|}
\hline
\textbf{Muon Selections} \\
\hline
Track must be fiducial to CMU, CMP, or CMX \\
$E_{EM} \le 2$ GeV \\
$E_{HAD} \le 6$ GeV \\
$E_{EM}+E_{HAD} \ge 0.1$ GeV for CMIO muons \\
$\mathrm{Iso}(R=0.4)/E_T \le 0.1$ \\
$\ge 3$ COT axial segments with $\ge 5$ hits \\
$\ge 2$ COT stereo segments with $\ge 5$ hits \\
$\ge 1$ Si hit \\
$z_0^\mathrm{track} \le 60$ cm \\
if $p_T > 300$ GeV, $n_{transitions} \ge 30$ \\
$d_0 < 0.02$ cm \\ 
$\Delta X_{CMU} \le 3$ cm \\
$\Delta X_{CMP} \le 5$ cm \\
$\Delta X_{CMX} \le 6$ cm \\
\hline
\hline
\textbf{Loose Muon Selections} \\
\hline
$p_T \ge 10$ GeV \\
\hline
\hline
\textbf{Tight Muon Selections} \\
\hline
Muon must be of type CMUP or CMX \\
$p_T \ge 20$ GeV \\
\hline
\end{tabular}
\caption{\label{tab:Muon_cuts}
Selections to identify muons.}
\end{table}

\begin{table}
\centering
\begin{tabular}{|c|}
\hline
Loose Central Electron Selections \\
\hline
$\eta_\mathrm{Detector} \le 1.1$ \\
Track must be fiducial to CES \\
$E_T \ge 12$ GeV \\
$E_{HAD}/E_{EM} \le 0.055 + 0.00045 \times E$ \\
$\mathrm{Iso}(R=0.4)/E_T \le 0.1$ \\
$p_T \ge 6$ GeV \\
$\ge 3$ COT axial segments with $\ge 5$ hits \\
$\ge 2$ COT stereo segments with $\ge 5$ hits \\
$z_0^\mathrm{track} \le 60$ cm \\
$E/p \le 2$ unless $p_T \ge 50$ GeV \\
$\chi^2_{CES} \le 10$ \\
$-3.0$ cm $ \le Q \times \Delta X_{CES} \le 1.5$ cm \\
$|\Delta Z_{CES}|<3$ cm \\
$L_{shr} \le 0.2$ \\
\hline
\end{tabular}
\caption{\label{tab:LCE_cuts}
Selections to identify loose central electrons}
\end{table}

\subsection{$W$ selection}

\label{sec:W_selection}

The $W$ boson selection requires a tight trigger lepton (defined above) 
 with $\Et>20$ GeV, $\met > 25$ GeV, and
transverse mass of lepton$+\met> 20$ GeV. In order to remove events where the $\met$ arises from a mismeasured lepton, the difference in $\phi$ between the
highest-energy lepton and the $\vec \met$ is required to be
greater than 0.5 radians.
The predicted and observed numbers of $W$ boson events selected from each trigger are summarized in Table~\ref{tab:EventCounts}.

\subsection{$Z$ Selection}
The $Z$ boson selection requires two oppositely-charged leptons: either two electrons or two muons.  In addition, the invariant mass of the two leptons must be between 76 GeV and 106 GeV.

Predictions are made in two dielectron categories: One electron from the $Z$ must be tight (Table~\ref{tab:TCE_cuts}), and the other may be tight or loose (Table~\ref{tab:LCE_cuts}).  The predicted and observed numbers of events in both of these categories are summarized separately in Table~\ref{tab:EventCounts}.

Predictions are also made in ten dimuon categories: One muon must be tight, and therefore either a CMUP or a CMX muon.  The other muon may be loose, and therefore may have any of the CMU, CMP, CMUP, CMX, or CMIO types.  The predicted and observed numbers of events in all of these categories are summarized in Table~\ref{tab:EventCounts}.

\begin{table}[!hb]
\centering
\begin{tabular}{|l|cc|}
\hline
Selection & Expected & Observed \\
\hline
W($e$, \met) & 2571230 & 2548108 \\
W($\mu_{CMUP}$, \met) & 1289610 & 1279001 \\
W($\mu_{CMX}$, \met) & 904569 & 895257 \\
Z($e$, $e$) & 156894 & 160251 \\
Z($e$, $e_{loose}$) & 25506 & 28896 \\
Z($\mu_{CMUP}$, $\mu_{CMU}$) & 8008 & 8391 \\
Z($\mu_{CMUP}$, $\mu_{CMP}$) & 9736 & 10433 \\
Z($\mu_{CMUP}$, $\mu_{CMUP}$) & 39620 & 36632 \\
Z($\mu_{CMUP}$, $\mu_{CMX}$) & 12893 & 13547 \\
Z($\mu_{CMUP}$, $\mu_{CMIO}$) & 9303 & 8489 \\
Z($\mu_{CMX}$, $\mu_{CMU}$) & 5860 & 6024 \\
Z($\mu_{CMX}$, $\mu_{CMP}$) & 6762 & 6863 \\
Z($\mu_{CMX}$, $\mu_{CMUP}$) & 14162 & 14467 \\
Z($\mu_{CMX}$, $\mu_{CMX}$) & 17245 & 17906 \\
Z($\mu_{CMX}$, $\mu_{CMIO}$) & 5852 & 5967 \\
\hline
\end{tabular}

\caption{\label{tab:EventCounts}
Event counts in $W$ and $Z$ boson samples, split up by categories of the leading and subleading leptons.}
\end{table}
\clearpage
}

\section{Soft lepton identification}

\label{sec:SoftLep}

The identification of low-$p_T$, or ``soft'', leptons is a main focus of this analysis.  Likelihood-based methods are used to identify soft electrons and muons.  The identification algorithms are described here, along with the methods used to validate them and evaluate their systematic uncertainties.

\subsection{Soft electrons}

Soft electrons are identified using a likelihood method trained on a signal sample from photon conversions and a background sample from other tracks with electron sources removed.

\subsubsection{Identification algorithm and candidate selections}

\ifthenelse{\boolean{Thesis}}
{
Every track in the event is a soft electron candidate, provided that it passes track quality selections and fiduciality selections:

\begin{itemize}\setlength{\itemsep}{0pt}
\item 20 axial and 20 stereo COT hits
\item At least 2 COT superlayers with 6 hits
\item Track extrapolates to CES, CPR, and calorimeter
\item Track $|\eta| < 1$.
\end{itemize}
}
{
A preselection is applied to all soft electron candidates requiring good track quality as well as 
track extrapolation to the CES, CPR, and calorimeter. Only tracks with $|\eta|<1$ are considered for the soft electron identification.
}

After this preselection, a likelihood-based calculator is used to identify electrons.  The likelihood calculator uses seven discriminating variables: 
the energy loss as the track traverses the tracking chamber, the electromagnetic and hadronic calorimeter energies, the energies deposited in the preradiator and the showermax detector, and the two-dimensional distance $(\Delta x, \Delta z)$ between the extrapolated position of the track and the shower in the CES. The calorimeter variables are calculated using a narrow, two-tower-wide section of the calorimeter.
\ifthenelse{\boolean{Thesis}}
{
Appendix~\ref{sec:appendix_se} contains more information on these discriminating variables.
}
{}

Some of the variables used in the soft electron identification are modeled very badly in the MC, and so the likelihood is trained on data without resorting to the simulation. For each of the above variables $x_i$, a fit is performed to the ratio of the distribution in the electron sample and the distribution in the non-electron background (``fake'') sample.  For each candidate, the value of each of these fit functions is multiplied together to get the final likelihood ($\mathcal L_{\rm electron}$):


$$Q = \prod_i \frac{P(x_i | {\rm real})}{P(x_i | {\rm fake})}, ~~~~~~~~~~
\mathcal{L}_{\rm electron} = \frac{Q}{1+Q}.
$$

The distribution of the likelihood in the real and fake samples is shown in Figure~\ref{fig:SE_compare}.  A candidate is identified as an electron if it passes the requirement $\mathcal{L}_{\rm electron} > 0.99$.

%
%
%

\subsubsection{Training samples and efficiency and misidentification rate measurements}
\label{se_training}

\begin{figure}
\begin{centering}
\includegraphics[width=0.45\textwidth]{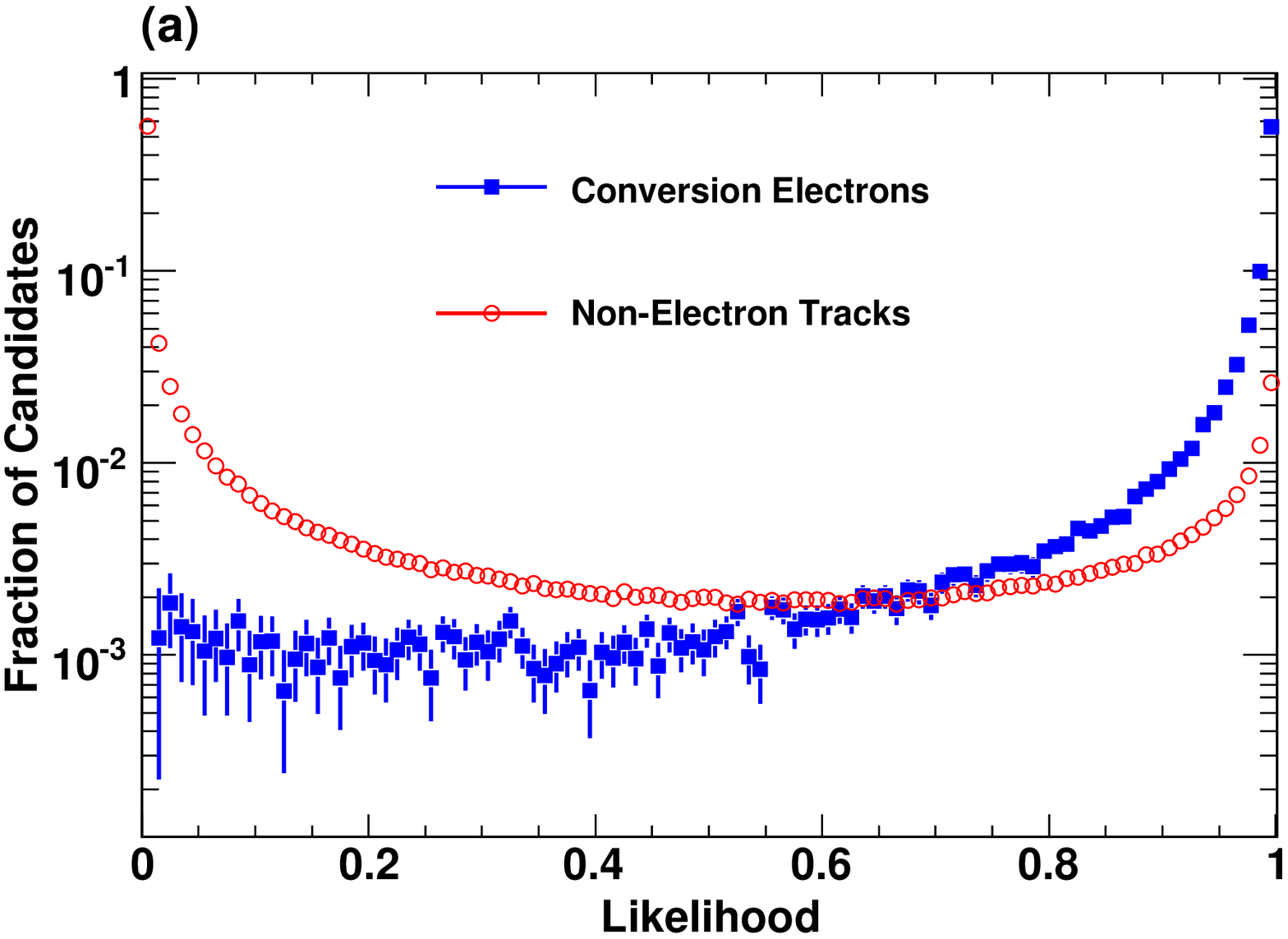}
\hfil
\includegraphics[width=0.45\textwidth]{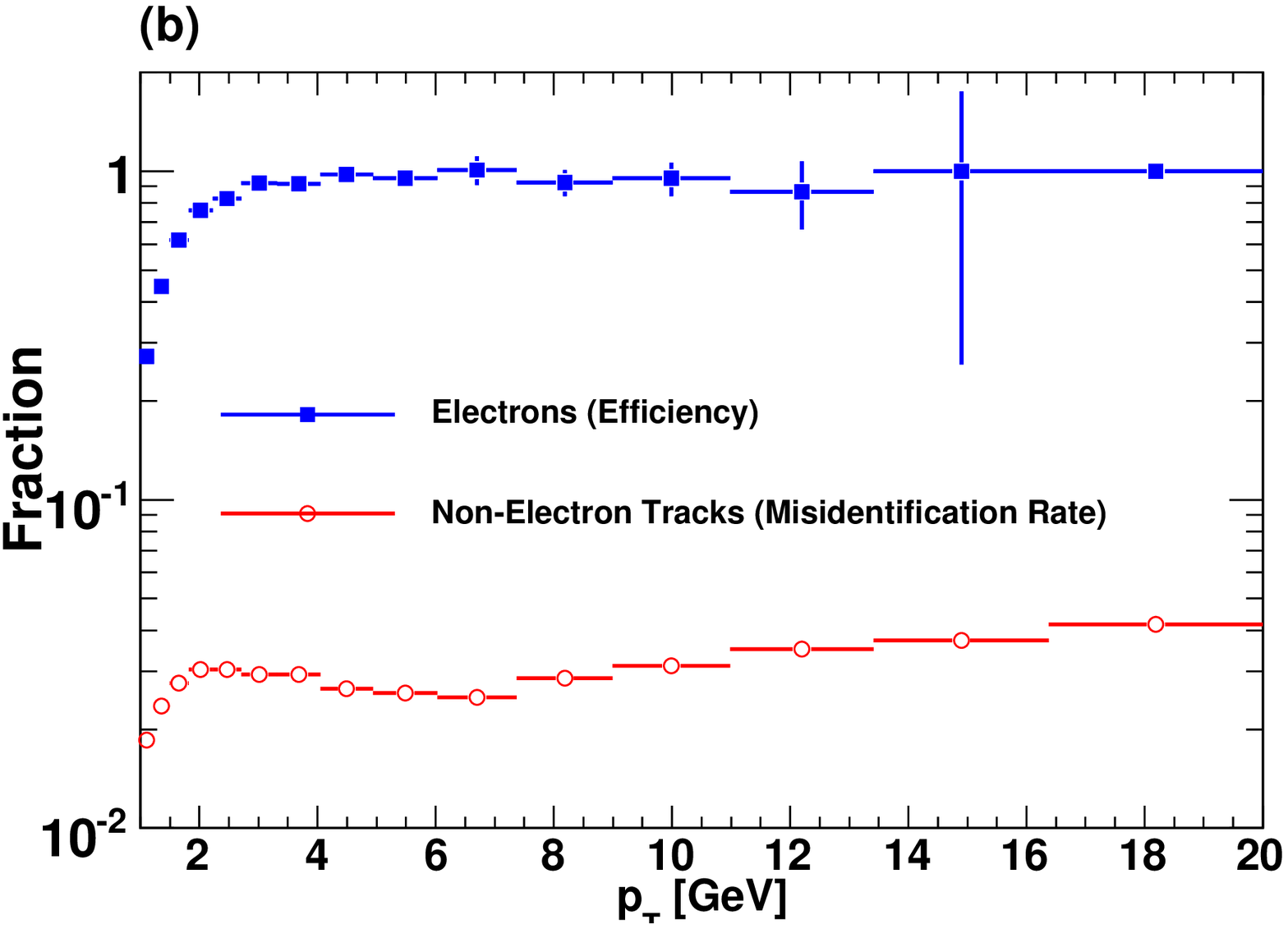}

\caption{\label{fig:SE_compare} 
(a) The likelihood distributions for electrons (closed squares) and non-electrons (open circles) after all preselection criteria. Only those candidates with a likelihood $> 0.99$ are identified as electrons.  (b) The efficiency as a function of $p_T$ for the identification of electrons (closed squares) and tracks misidentified as electrons (open circles) after the likelihood selection.
}
\end{centering}
\end{figure}

Photon conversions are used as a pure sample of electrons to train the likelihood function.
\ifthenelse{\boolean{Thesis}}
{
The $8$ GeV electron trigger (see Appendix~\ref{sec:appendix_SE_samples}) is used to obtain a pure sample of conversions by requiring
that there be two tracks with opposite sign having $|\delta|<0.2$ cm, $\Delta \cot(\theta) < 0.1$, and $R_{Conv}>8$ cm. Figure~\ref{fig:conv_vars}a illustrates
these variables.

\begin{figure} \centering
\includegraphics[width=2in]{plots/conv_vars.eps}
\hfil
\includegraphics[width=0.45\textwidth]{plots/DelCot.eps}
\caption{\label{fig:conv_vars}
Variables used to identify photon conversions to electron-positron pairs.
On the left are the variables defined in the plane transverse to the beam. The beam position is denoted by an ``x''. $R$ is the distance
between the beam position and the point at which the two tracks are tangential or parallel to each other and $\delta$ is the distance between
the two tracks at that point.
On the right is the distribution of $\Delta\cot(\theta)$, where
$\theta$ is the polar angle of the track in the $r-z$ plane. A fit is performed to find the signal (solid) and background (dashed) to estimate the sample composition under the peak.}
\end{figure}

After these selections, a fit is performed to the $\Delta\cot(\theta)$ distribution, shown in Figure~\ref{fig:conv_vars}b,
to determine the non-conversion background under the peak.  The sideband of the distribution ($0.06 < |\Delta\cot(\theta)| < 0.1$) is used to subtract out this background.

The likelihood distributions for the electron sample and for the non-electron background sample are shown in Figure~\ref{fig:SE_compare} (left).


Since the higher-$p_T$ track of the conversion pair will be trigger-biased, the likelihood is trained using the softer track.  A conversion pair is not used if the hard track extrapolates to the same calorimeter towers that are used for the soft track, since those conversions have a very different $E_{em}/p$ distribution.
}
{
In events selected using an 8 GeV electron trigger, pairs of tracks are found that correspond to a photon converting into $e^+e^-$~\cite{MyThesis}.  In order to avoid any bias from the trigger, the lower-momentum track of the conversion pair is used to train the likelihood.
}

Events from the 18 GeV muon trigger are used to select a sample of non-electron tracks with which to train the likelihood function\ifthenelse{\boolean{Thesis}}{ (See Appendix~\ref{sec:appendix_SE_samples})}{}.  All tracks in the events that, along with another track, form a possible photon conversion are removed from the training sample.  To reduce the bias from using a muon-triggered sample, any track that is within $\Delta R <0.7$ of an identified muon is also ignored.  In addition, to reduce the contamination from real electrons, any event that contains an identified heavy quark decay or an identified high-$p_T$ electron is ignored.



The efficiency and fake rate are calculated in these training samples as functions of $p_T$, $\eta$, and track isolation.  The same sample used for training is also used to measure the efficiency, due to the larger backgrounds present in other independent samples.  The separation in identification rate between electrons and non-electrons after the likelihood selection is shown in Figure~\ref{fig:SE_compare} (right). The efficiency in terms of $p_T$ and $\eta$, after the track and CES shower have been identified, is shown in Table~\ref{tab:se_eff}.

\begin{table}[htb]
\caption{\label{tab:se_eff} Efficiency to identify soft (2 GeV $<p_T<$ 20 GeV) electrons as a function of candidate $p_T$ and $\eta$.}
\begin{center}
\begin{tabular}{cccccccc}
\hline
\hline
$p_T$ range (GeV) & $[2, 2.5]$ & $[2.5, 3]$ & $[3, 6]$ & $[6, 12]$ & $[12, 20]$ \\
$0<|\eta|<0.2$ & 0.90 & 0.99 & 0.99 & 0.90 & 0.99 \\
$0.2<|\eta|<0.6$ & 0.94 & 0.99 & 0.99 & 0.99 & 0.99 \\
$0.6<|\eta|<1$ & 0.95 & 0.99 & 0.99 & 0.99 & 0.99 \\
\hline
\hline
\end{tabular}
\end{center}
\end{table}


This identification
rate is applied as a weight to each candidate track in the MC to find the predicted number of identified electrons.

\ifthenelse{\boolean{Thesis}}
{
Some representative fits are shown in Figure~\ref{fig:SE_param_fits}. To show that this scheme takes into account any correlations between these kinematic variables, the results of applying these parametrizations to the training samples are shown in Figure~\ref{fig:se_param}.

\begin{figure}
\begin{center}
\includegraphics[width=0.45\textwidth]{plots/se_param_fit_real.eps}
\hfil
\includegraphics[width=0.45\textwidth]{plots/se_param_fit_fake.eps}

\caption{\label{fig:SE_param_fits}Examples of fits to the soft electron efficiency (left) and fake rate (right) as a function of track isolation in a particular $p_T$ and $\eta$ bin.}
\end{center}
\end{figure}

\begin{figure}
\begin{center}
\includegraphics[width=0.49\textwidth]{plots/SE_param_real_pt.eps}
\hfil
\includegraphics[width=0.49\textwidth]{plots/se_param_fake_pt.eps}

\includegraphics[width=0.49\textwidth]{plots/SE_param_real_eta.eps}
\hfil
\includegraphics[width=0.49\textwidth]{plots/se_param_fake_eta.eps}

\includegraphics[width=0.49\textwidth]{plots/SE_param_real_iso.eps}
\hfil
\includegraphics[width=0.49\textwidth]{plots/se_param_fake_iso.eps}

\caption{\label{fig:se_param} 
A comparison of the measured and predicted efficiencies (left) and fake rates (right). The kinematic variables shown, from top to bottom, 
transverse momentum $p_T$, pseudo-rapidity $\eta$, and track isolation are those used to characterize the response of the soft electron
algorithm.
}
\end{center}
\end{figure}
}
{}

\subsubsection{Validation and systematic uncertainty determination}
\label{sec:se_validation}

The efficiency and fake rate parametrizations are checked on a data set triggered on jets having $E_T>50$ GeV\ifthenelse{\boolean{Thesis}}{ (See Appendix~\ref{sec:appendix_SE_samples})}{}. The parametrizations use the $p_T$, $\eta$ and isolation of candidates in order to account for any kinematic differences between the training sample and the validation sample.  
First, the same electron removal that was used for the fake training sample (Section~\ref{se_training}) is applied to the tracks in the jet sample.  The likelihood distribution of all candidate tracks in the jet sample is then fit to templates from the real and fake likelihood training samples to obtain the fraction of real and fake electrons in the jet sample.  The jet sample is found to consist of 2.5\% real electrons, mostly coming from photon conversions from which only one electron was reconstructed.  The predicted identification rate is then checked for agreement with the measured identification rate.

The disagreement between the calculated and observed identification rates is measured 
to be 1.6\%.  However, we observe larger disagreement in the shapes of the calculated and observed distributions in $p_T$ and $\eta$\ifthenelse{\boolean{Thesis}}{, as seen in Figure~\ref{fig:SE_validation}}{}. We assign a systematic uncertainty of 15\%, which is sufficient to cover the observed disagreement~\cite{MyThesis}.  This systematic uncertainty is applied separately to the electron identification and misidentification rates.

\ifthenelse{\boolean{Thesis}}
{
\begin{figure}

\includegraphics[width=0.45\textwidth]{plots/se_valid_pt.eps}
\hfil
\includegraphics[width=0.45\textwidth]{plots/se_valid_eta.eps}

\caption{\label{fig:SE_validation} 
Predicted and observed soft electron misidentification rates obtained from a QCD (jet) sample. On the left are the identification rates as function of $p_T$, and on the right the the identification rates as functions of $\eta$.}
\end{figure}
}
{}

\subsection{Soft muons}
\label{sec:softmu}

\ifthenelse{\boolean{Thesis}}
{
The soft muon identification algorithm described in Ref.~\cite{SoftMuonDescription} has been ported from the {\sc TopNtuple} framework to 
the {\sc Stntuple} framework with minimal modifications. The efficiency of the ported identification software is measured using 
reconstructed $J/\psi\rightarrow\mu^{+}\mu^{-}$ decays to obtain pure muon samples. The misidentification rates of pions and kaons
are measured in $D^{*+}\rightarrow D^{0}\pi^{+}$ decays where $D^{0}$ decays as $D^{0}\rightarrow K^{-}\pi^{+}$. Similarly, the
proton misidentification rate is measured in $\Lambda^{0}\rightarrow p\pi$.  First, the algorithm is summarized.
}
{
Soft muons are identified using a method similar to that described in Ref.~\cite{SMtop}. The inputs to the algorithm are derived from a sample of muons arising from $J/\psi$ decays in a muon calibration dataset.
}

\subsubsection{Identification algorithm and candidate selections}
\label{sec:softmu_idselect}

The soft muon identification algorithm relies on matching tracks identified in the COT to track segments reconstructed in the muon 
chambers (muon stubs). Matching is done in the extrapolated position along the muon chamber drift direction $(x)$, the longitudinal coordinate along the
chamber wires $(z)$ when available, and the difference in slope between the extrapolated COT track and the reconstructed muon chamber track segment $(\phi_{L})$. 
Tracks are paired with muon chamber track segments based on the best match in $x$ for those track segments within $50$ cm of an extrapolated COT track.

\ifthenelse{\boolean{Thesis}}
{
The list of all such track-stub matching variables is:
\begin{itemize} \setlength{\itemsep}{0pt}
\item CMU $dx$,
\item CMU $dz$,
\item CMU $d\phi$,
\item CMP $dx$,
\item CMP $d\phi$,
\item CMX $dx$,
\item CMX $dz$,
\item CMX $d\phi$.
\end{itemize}
Each of these variables can only be used if a stub exists in the corresponding system, i.e. CMP $dx$ and $d\phi$ only have values if there
is a CMP stub.
}
{}

\ifthenelse{\boolean{Thesis}}
{
Table~\ref{tab:softmu_candidate} lists the requirements placed on soft muon candidate tracks.
The selections on the number of COT hits reduce the background from poorly measured tracks, 
while the selection on impact parameter removes some of the pion and kaon decay-in-flight background. 
Note that the fiducial definition described in the last bullet point differs from the one used in Ref.~\cite{SoftMuonDescription}. In that
algorithm, candidates were declared fiducial if they extrapolated to within $3\sigma_{MS}$ outside of the physical chamber 
boundary, where $\sigma_{MS}$ is the width of the multiple scattering distribution for a given $p_T$. This change was made
due to the unavailability of the extrapolated track-to-chamber boundary distance in the {\sc Stntuple} format.
\begin{table}
\centering
\begin{tabular}{|c|}
\hline
\textbf{Soft Muon Candidate Selections}\\
\hline
$N(\mathrm{COT})\geq48$ \\
$N(\mathrm{COT~Axial})\geq24$ \\
$N(\mathrm{COT~Stereo})\geq24$ \\   
$|d_{0}|<0.3$ cm, where $d_{0}$ is the impact parameter with respect to the beamline \\
$|z_{0}|<60$ cm \\
The track must extrapolate to within the physical boundary of a muon chamber.\\
\hline
\end{tabular}
\caption{\label{tab:softmu_candidate}
Soft muon candidate selection criteria.
}
\end{table}
}
{
Soft muon candidates are required to extrapolate to within the physical boundaries of a muon chamber, 
have good track quality, have at least one hit in the SVX, $|d_{0}|<0.3$ cm where $d_{0}$ is the impact parameter with respect to the beamline, and 
$z_{0}<60$ cm where $z_{0}$ is the $z$ position of the track at the interaction point.
}

\ifthenelse{\boolean{Thesis}}
{
A $\chi^{2}$ is built from the track-to-stub matching variables $x_i$ described above:
\begin{equation}
\label{eq:softmu_Chi}
\chi^2=\sum^{n}_{i}\frac{(x_i-\mu_i)^{2}}{\sigma_{i}^{2}}=\sum^{n}_{i}y_{i}^{2},
\end{equation}
where $\mu_i$ and $\sigma_{i}^{2}$ are the expected mean and variance of the distribution of $x_i$. 
The final scaled $\chi^2$ is calculated by normalizing $\chi^2$:
\begin{equation}
\label{eq:softmu_Q}
\mathcal{Q}_{\rm muon}=\frac{\chi^2-n}{\sigma(\chi^2)},
\end{equation}
where $\sigma(\chi^2)$ is the expected standard deviation of $\chi^2$. 

The track-to-stub matching variance functions, $\sigma_i$ in Eq.~\ref{eq:softmu_Q}, are copied from the {\sc TopNtuple} code directly
into the {\sc Stntuple} port with no modifications. The $\sigma(Q)$ term in the denominator of Eq.~\ref{eq:softmu_L} is decomposed
as in Ref.~\cite{SoftMuonDescription}
\begin{equation}
\sigma^{2}(Q)=2n+\sum_{i,j}\rho(y_{i}^{2},y_{j}^{2}),
\end{equation}
and the $\rho$'s are taken from the {\sc TopNtuple} code.
}
{
A $\chi^{2}$ is built from the track-to-stub matching variables $x_i$ described above ($dx$, $dz$, and $d\phi_{L}$).  This $\chi^2$ is normalized to have mean 0 and variance 1 for real muons, independent of the number $n$ of track-stub matching variables $x_i$ used:
$$ \chi^2 = \sum_{i}\frac{(x_i-\mu_i)^{2}}{\sigma_{i}^{2}}, ~~~~~~~~~~ \mathcal{Q}_{\rm muon} = \frac{\chi^2-n}{\sigma(\chi^2)}, $$
where $\mu_i$ and $\sigma_{i}^{2}$ are the expected mean and variance of the distribution of $x_i$, and $\sigma(\chi^2)$ is the expected standard deviation of $\chi^2$.
}

In the final selection, we require that all identified soft muons must have a track segment in each muon chamber
to which the track extrapolates and that $|\mathcal{Q}_{\rm muon}|<3.5$ (see Fig.~\ref{fig:smu_likeComp}).
\ifthenelse{\boolean{Thesis}}
{
For example, if a track should cross the physical
volume of both the CMU and CMP detectors, there must be stubs in both detectors for it to be identified as a soft muon.
}
{}

\subsubsection{Efficiency and misidentification rate measurements}

\ifthenelse{\boolean{Thesis}}
{

The efficiency of the soft muon identification is measured using a pure sample of muons selected from $J/\psi \to \mu\mu$ decays.  These events are obtained using an online trigger requiring the presence
of a muon with $p_T>8$ GeV (See Appendix~\ref{sec:appendix_SM_samples}), and the $J/\psi$ is reconstructed by requiring
that the trigger muon make a vertex with another track of opposite charge that has associated muon chamber hits.  All requirements listed in Sec.~\ref{sec:softmu_idselect} are applied to both tracks.
The $J/\Psi$ candidate mass is required to be consistent with $2.9<m(\mu\mu)<3.3$ GeV, and signal and sideband regions are defined as follows:
\begin{itemize}\setlength{\itemsep}{0pt}
\item Left Sideband: $2.94<m(\mu\mu)<3.0$ GeV,
\item Signal Region: $3.03<m(\mu\mu)<3.15$ GeV,
\item Right Sideband: $3.18<m(\mu\mu)<3.24$ GeV,
\end{itemize} 
The signal and sideband yields are fit in bins of $p_T$ of the non-triggered leg and the fits are used to subtract out the background under the mass peak.  These fits are shown in Fig.~\ref{fig:mjpsiFit_pTbins}.

\begin{figure}[htb]
\begin{center}
\includegraphics[width=0.9\textwidth]{plots/mjpsiFit_pTbins}
\caption{\label{fig:mjpsiFit_pTbins}
  Results of the $J/\psi$ mass fits in bins of the $p_{T}$ of the softer, 
 not-triggered, candidate leg of the $J/\psi$. 
}
\end{center}
\end{figure}

The misidentification rates of pions and kaons
are measured in $D^{*+}\rightarrow D^{0}\pi^{+}$ decays where $D^{0}$ decays as $D^{0}\rightarrow K^{-}\pi^{+}$.
These events are obtained from the two track trigger (See Appendix~\ref{sec:appendix_SM_samples}) and are reconstructed with the following selections:

\begin{itemize}\setlength{\itemsep}{0pt}
\item the $K$ must have opposite charge to each of the two $\pi$'s,
\item $|z|\leq5$ cm between any two tracks,
\item the soft pion from the $D^{*}\rightarrow D^{0}$ decays must have $p_T\geq500$ MeV,
\item the kaon and pion from the $D^{0}$ decay must have $p_T\geq2$ GeV,
\item the kaon and pion from the $D^{0}$ decay must have $|d_{0}|\leq0.2$ cm,
\item $m(K\pi)-m(D^{0})\leq30$ MeV where $m(D^{0})$ is the nominal $D^{0}$ mass,
\item $p_T(D^{0})\geq5$ GeV,
\item the impact parameter significance of the $D^{0}$ is required to be $d_{0}/\sigma(d_{0})\geq2$,
\item $p_T(D^{*})\geq6$ GeV,
\item $\Delta m=m(D^{*})-m(D^{0})\leq170$ MeV.
\item $\chi^{2}\leq100$ where $\chi^{2}$ is from the vertex fit.
\end{itemize} 

The $\pi$ and $K$ from the $D^{0}$ are required to form a vertex while the slow $\pi$ from the $D^{*}$ is attached to the primary vertex. The
$D^{0}$ vertex is required to point back to the primary vertex. Signal and sideband regions are defined as follows:

\begin{itemize}\setlength{\itemsep}{0pt}
\item Left Sideband: $0.1396<\Delta m<0.141$ GeV,
\item Signal Region: $0.14242<\Delta m<0.148421$ GeV,
\item Right Sideband: $0.152<\Delta m<0.1625$ GeV,
\end{itemize}

The signal and sideband yields are fit in bins of $p_T$ of the $\pi$ or $K$ from the $D^{0}$. These fits are shown in Fig.~\ref{fig:dstFit_pi_pTbins}.

\begin{figure}[htb]
\begin{center}
\includegraphics[width=0.9\textwidth]{plots/dstFits_pi_pTbins}
\caption{\label{fig:dstFit_pi_pTbins}
  Results of the $D^{*}$ mass fits in bins of the $p_{T}$ of candidate $\pi$'s coming from $D^{0}\rightarrow K\pi$.
}
\end{center}
\end{figure}

The misidentification rate of protons is measured using a sample of protons obtained from $\Lambda\rightarrow p\pi$ decays.  These events are taken from the two track trigger (See Appendix~\ref{sec:appendix_SM_samples}).  The selections are as follows:
\begin{itemize}\setlength{\itemsep}{0pt}
\item the two tracks must pass the selections in Sec.~\ref{sec:softmu_idselect},
\item the two tracks are required to have opposite charge and fit to a vertex,
\item $|\Delta z|\leq2$ cm between the two tracks,
\item the $\chi^{2}$ of the vertex fit is required to be $<10$,
\item the decay length significance of the vertex is required to be $L_{xy}/\sigma(L_{xy})\leq20$,
\item the $\Lambda$ impact parameter is required to be $|d_{0}|<0.02$ cm,
\item $1.0<m(\Lambda)<1.16$ GeV.
\end{itemize} 

Signal and Sideband regions are defined as follows:

\begin{itemize}\setlength{\itemsep}{0pt}
\item Left Sideband: $1.101<m(\Lambda)<1.106$ GeV,
\item Signal Region: $1.111<m(\Lambda)<1.121$ GeV,
\item Right Sideband: $1.126<m(\Lambda)<1.131$ GeV,
\end{itemize}

The signal and sideband yields are fit in bins of proton $p_T$. These fits are shown in Fig.~\ref{fig:lamFit_pTbins}.

\begin{figure}[htb]
\begin{center}
\includegraphics[width=0.9\textwidth]{plots/lamFit_pTbins}
\caption{\label{fig:lamFit_pTbins}
  Results of the $\Lambda$ mass fits in bins of the $p_{T}$ of candidate $p$'s.
}
\end{center}
\end{figure}

Figure~\ref{fig:LikeComp_allTypes} shows the distribution of the scaled $\chi^2$ $\mathcal{Q}_{\rm muon}$ for $\mu$, $\pi$, $K$, and $p$ obtained from the signal regions in the samples described above. The muon sample peaks more strongly at small $\mathcal{Q}_{\rm muon}$, as expected. The final selection, as described in Sec.~\ref{sec:softmu_idselect}, is $|\mathcal{Q}_{\rm muon}|<3.5$.

\begin{figure}[htb]
	\begin{center}
		\includegraphics[width=0.8\textwidth]{plots/LikeComp_softMu_allTypes}
			\caption{\label{fig:LikeComp_allTypes}
			  A comparison of the soft muon scaled $\chi^2$ distributions for $\mu$, $\pi$, $K$, and $p$. 
			}
			\end{center}
\end{figure}

The technique described in Ref.~\cite{SoftMuMistag} is used to obtain the efficiency and fake rates. The identification
rate is determined as,
\begin{equation}
R_{S}=\frac{R_{M}-R_{B}\cdot f_{B}}{1-f_{B}},
\end{equation}
where $R_{M}$ and $R_{B}$ are the identification rates measured in the signal and sideband regions, respectively, and $f_{B}$ is the
background fraction in the signal window. For fake rates, the measured misidentification rate, $R_{M}$, can be written in terms of the decay-in-flight rate as,

\begin{equation}
R_{M}=f_{M}\cdot R_{DIF}+(1-f_{M})\cdot R_{PT},
\end{equation}

where $f_{M}$ is the decay-in-flight fraction and $R_{DIF}$ and $R_{PT}$ are the identification rates for decay-in-flight and punch-through, respectively. The $R_{DIF}$ is used as presented in 
Ref.~\cite{SoftMuMistag}.

The identification efficiency is defined as $N(\mathrm{identified})/N(\mathrm{candidates})$, where the candidate requirements are shown in Table~
\ref{tab:softmu_candidate}. The efficiencies for each particle type are plotted in Fig.~\ref{fig:EffComp_softMuTag} as a function of $p_T$.
Note that for all particles except for muons, the ``efficiency'' is actually the rate that the particle is \emph{misidentified} as a muon.
Strong separation is observed between $\mu$ and backgrounds.  
\begin{figure}[htb]
	\begin{center}
		\includegraphics[width=0.8\textwidth]{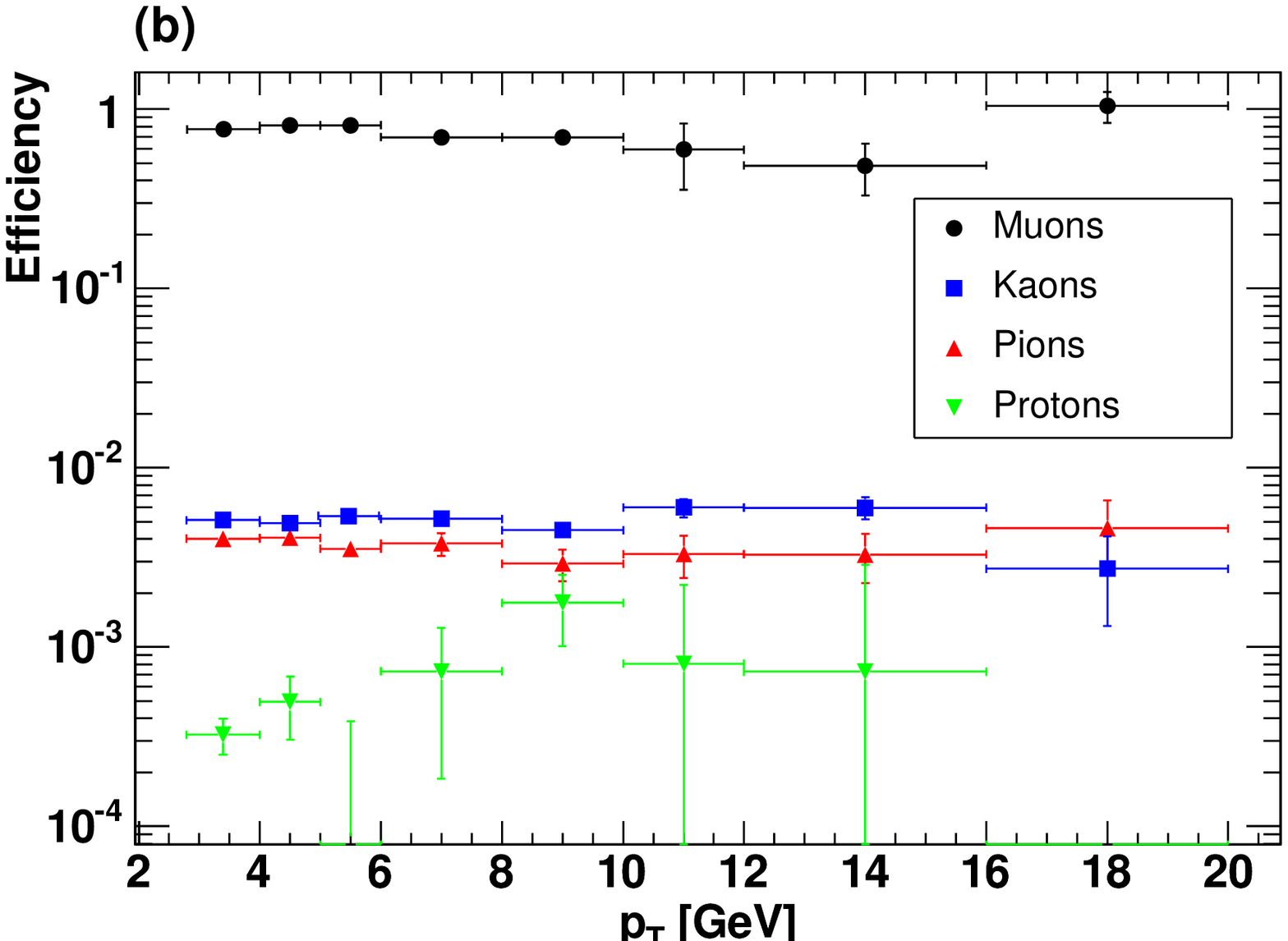}
			\caption{\label{fig:EffComp_softMuTag}
			  Identification efficiency as a function of $p_T$ for $\mu$, $\pi$, $K$, and $p$. For the case of the $\mu$, this is
			  the rate at which real muons are identified. For the other species, it is the rate that the particle is
			  misidentified as a muon. 			  
			}
			\end{center}
\end{figure}

An efficiency matrix is created in bins of $p_T$ and $\eta$ using the $J/\psi$ sample. Because the sample is limited in statistics for $p_T>12$ GeV, empty bins are filled in using interpolation between the low-$p_T$ muons from $J/\Psi$ decays and higher-$p_T$ muons from $Z$ decays.  The soft
muon identification is applied to $Z$ events so that the region between the $J/\psi$ and $Z$ $p_T$ may be correctly fitted.  Figure~\ref{fig:pTeffFit_eta0}
shows an example of these fits for candidates with $|\eta|<0.15$.
The final $\mu$ efficiency matrix is shown in Fig.~\ref{fig:TRMatrix_mu}.

\begin{figure}[htb]
	\begin{center}
		\includegraphics[width=0.8\textwidth]{plots/pTEff_etaBand0_muo.eps}
			\caption{\label{fig:pTeffFit_eta0}
			  Fit for soft muon efficiency function over $J/\psi$ and $Z$ events in the $\eta$ range $|\eta|<0.15$.
			}
			\end{center}
\end{figure}

%
%
%
\begin{figure}[htb]
	\begin{center}
		\includegraphics[width=0.75\textwidth]{plots/TRMatrix_mu}
			\caption{\label{fig:TRMatrix_mu}
			  Soft muon ID rate matrix in bins of $p_{T}$ and $\eta$.
			}
			\end{center}
\end{figure}

For the corresponding binned misidentification matrix, the misidentification rate is measured in each of the three background samples.  The $\pi$, $K$, and $p$ matrices are then combined in the proportion found in $W$ boson decays as presented in Table 3 of Ref.~\cite{SoftMuMistag}.  These weights are $f(\pi)=0.719$, $f(K)=0.156$, and $f(p)=0.125$. 

}
{
The efficiency of the soft muon identification is measured using a pure sample of muons obtained from $J/\psi \to \mu\mu$ decays.  These events are obtained using an online trigger requiring the presence
of a muon with $p_T>8$ GeV. The $J/\psi$ is reconstructed by requiring
that the trigger muon make a vertex with another track of opposite charge that has associated muon chamber hits.  All track requirements listed in Sec.~\ref{sec:softmu_idselect} are applied to both tracks.
The $J/\Psi$ candidate mass is required to satisfy $3.03<m(\mu\mu)<3.15$ GeV, and the sidebands of the mass distribution are used to evaluate the background under the mass peak.

The misidentification rates of pions and kaons
are measured in $D^{*+}\rightarrow D^{0}\pi^{+}$ decays where the $D^{0}$ decays as $D^{0}\rightarrow K^{-}\pi^{+}$.
These events are obtained from a trigger that requires the presence of a vertex containing two tracks and are reconstructed requiring masses $1.835<m(K\pi)<1.895$ GeV and $m(D^*)-m(D^0)<170$ MeV.
The sidebands of the $m(D^*)-m(D^0)$ distribution are used to evaluate the background under the mass peak.

The misidentification rate of protons is measured using a sample of protons obtained from $\Lambda\rightarrow p\pi$ decays.  These events are taken from the same dataset as that from which the
$D^{*}$ sample is obtained. The reconstructed $\Lambda$ mass is required to satisfy $1.111 < m(p\pi) < 1.121$ GeV. The sidebands of the mass distribution are used to evaluate the background under the mass peak.

Figure~\ref{fig:smu_likeComp} (left) shows the distribution of muon scaled $\chi^2$, $\mathcal{Q}_{\rm muon}$, using the samples described above.
Good separation is obtained between muons and other particle species.

\begin{figure}[htb]
   \begin{center}
    \ifthenelse{\boolean{Thesis}}{\includegraphics[width=0.45\textwidth]{plots/LikeComp_softMu_allTypes}}
    {\includegraphics[width=0.45\textwidth]{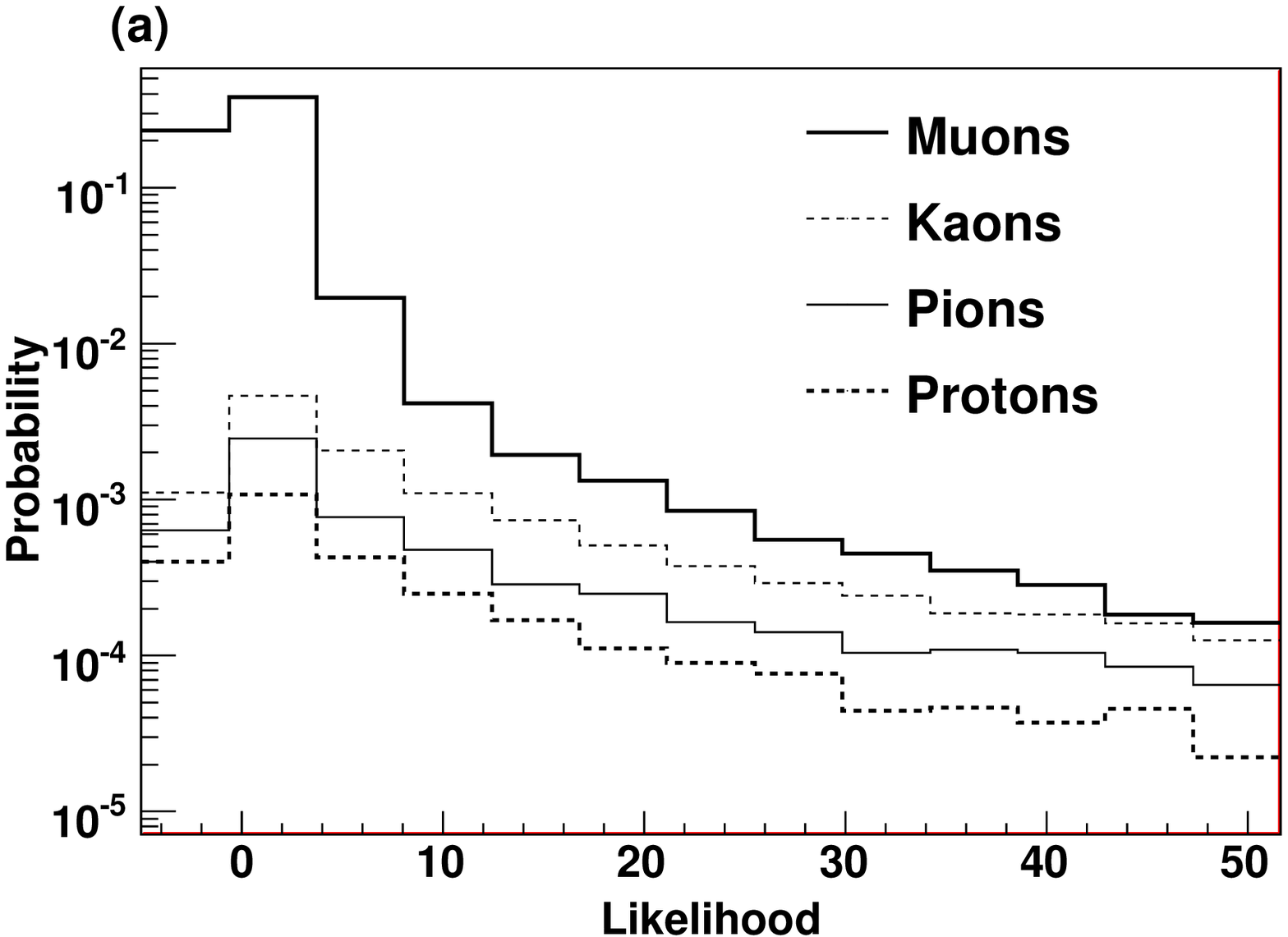}}
    \hfil
    \includegraphics[width=0.45\textwidth]{plots/EffComp_softMuTag}
    \caption{\label{fig:smu_likeComp}
    (a) The distribution of soft muon scaled $\chi^2$, $\mathcal{Q}_{\rm muon}$, for muons, pions, kaons, and protons after all preselection selections. Only those candidates with $|\mathcal{Q}_{\rm muon}| < 3.5$ are identified as muons.  (b) The muon identification rate (circles) and misidentification rates for pions (triangles), kaons (squares), and protons (triangles) after the scaled $\chi^2$ selection.}
    \end{center}
\end{figure}

\begin{table*}[htb]
\caption{\label{tab:sm_eff} Efficiency to identify soft muons as a function of candidate $p_T$ and $\eta$.}
\begin{center}
\begin{tabular}{ccccccccc}
\hline
\hline
$p_T$ range (GeV) & $[3, 4]$ & $[4, 5]$ & $[5, 6]$ & $[6, 8]$ & $[8,10]$ & $[10,12]$ & $[12,16]$ & $[16,20]$ \\
$-1.5<\eta<-0.7$ & 0.739 & 0.626 & 0.567 & 0.419 & 0.342 & 0.127 & 0.237 & 0.174 \\
$-0.7<\eta<-0.55$ & 0.593 & 0.556 & 0.581 & 0.480 & 0.438 & 0.299 & 0.356 & 0.344 \\
$-0.55<\eta<-0.45$ & 0.749 & 0.788 & 0.883 & 0.751 & 0.783 & 0.608 & 0.644 & 0.659 \\
$-0.45<\eta<-0.15$ & 0.816 & 0.901 & 0.898 & 0.782 & 0.821 & 0.701 & 0.570 & 0.659 \\
$-0.15<\eta<0.15$ & 0.777 & 0.796 & 0.784 & 0.667 & 0.657 & 0.525 & 0.424 & 0.616 \\
$0.15<\eta<0.45$ & 0.832 & 0.918 & 0.913 & 0.799 & 0.815 & 0.698 & 0.568 & 0.659 \\
$0.45<\eta<0.55$ & 0.768 & 0.782 & 0.840 & 0.741 & 0.582 & 0.758 & 0.529 & 0.659 \\
$0.55<\eta<0.7$ & 0.625 & 0.573 & 0.556 & 0.461 & 0.450 & 0.409 & 0.237 & 0.256 \\
$0.7<\eta<1.5$ & 0.750 & 0.617 & 0.593 & 0.428 & 0.327 & 0.146 & 0.173 & 0.174 \\
\hline
\hline
\end{tabular}
\end{center}
\end{table*}

An efficiency matrix is created in bins of $p_T$ and $\eta$ using the $J/\psi$ sample. Because the sample is limited in statistics for $p_T>12$ GeV, empty bins are filled in using interpolation between the low-$p_T$ muons from $J/\Psi$ decays and higher-$p_T$ muons from $Z$ decays.  The soft
muon identification is applied to $Z$ events so that the region between the $J/\psi$ and $Z$ $p_T$ may be correctly fitted.  Note that Figure~\ref{fig:smu_likeComp} shows the observed results in these low-statistics bins, while Table~\ref{tab:sm_eff} shows the interpolated efficiencies.

For the corresponding binned misidentification matrix, the misidentification rate is measured in each of the three background samples.  The $\pi$, $K$, and $p$ matrices are then combined in the proportion found in $W$ boson decays.  These relative proportions are found to be $f(\pi)=0.719$, $f(K)=0.156$, and $f(p)=0.125$.

The efficiency and fake rate as a function of $p_T$ is shown in Figure~\ref{fig:smu_likeComp}.  The efficiency in terms of $p_T$ and $\eta$ is tabulated in Table~\ref{tab:sm_eff}.  These identification rates are measured after the track and muon hit(s) have been identified.

This identification
rate is applied as a weight to each candidate track in the MC to find the predicted number of identified electrons.

}

\subsubsection{Soft muon systematic uncertainty determination}
\label{sec:sm_syst}

\ifthenelse{\boolean{Thesis}}
{

Separate systematic uncertainties are estimated for the true muon identification efficiency and the misidentification rate. 
The sideband subtraction technique used to obtain the muon efficiency matrix introduces uncertainties arising from the statistics of the 
$J/\psi$ sample. These uncertainties vary with $p_T$ and $\eta$. 

In addition, the maximum variation in efficiency of $8\%$ arising from the difference between isolated and non-isolated candidates (See Figure~\ref{fig:SMisoeff}) is used
as an uncertainty representing the maximum possible difference between the $J/\psi$ sample environment and the $W/Z$ environment. This is added
in quadrature to the statistical uncertainty arising from the sideband subtraction method (2\% - 70\%, depending on the bin) to obtain the final muon efficiency uncertainty.


\begin{figure}[htb]
	\begin{center}
		\includegraphics[width=0.75\textwidth]{plots/SMisoeff}
			\caption{\label{fig:SMisoeff}
			  Soft muon identification efficiency as a function of the fractional isolation of the
			  muon obtained from the $J/\psi$ sample.
			}
			\end{center}
\end{figure}

The misidentification systematic uncertainty is obtained by selecting muon-free regions in JET samples and taking the difference between
observed and predicted soft muon misidentification rates. The JET sample selections are as follows:
\begin{itemize}\setlength{\itemsep}{0pt}
\item At least 3 jets with $E_T>15$ GeV and $|\eta|<2.0$,
\item Reject jets with positive {\sc SECVTX} tag or negative {\sc SECVTX} tag having $m(SV)>0.3$ GeV,
\item Reject candidate tracks in jets having $d_{0}/\sigma(d_{0})>2$.
\end{itemize}

Figure~\ref{fig:JET100_frValid_pTtrk} shows the predicted and observed identification rate in JET100.

\begin{figure}[htb]
	\begin{center}
		\includegraphics[width=0.75\textwidth]{plots/JET100_frValid_pTtrk}
			\caption{\label{fig:JET100_frValid_pTtrk}
			  Observed and predicted soft muon rate in the JET100 sample with the selections as described in the text
			  as a function of $p_T$.
			}
			\end{center}

\end{figure}

The differences between the predicted and observed number of events are shown in Table~\ref{tab:softmu_JETComp}. Twice the largest
error is used as the systematic uncertainty on the misidentification rate.

\begin{table}
\centering
\begin{tabular}{|l||c||c||c|}
\hline
\textbf{Sample} & \textbf{Identified} & \textbf{Predicted} & \textbf{Uncertainty} \\
\hline
JET50 & 517 & 505 & $2.3\%$ \\
JET100 & 2331 & 2220 & $4.8\%$\\
\hline
\end{tabular}
\caption{\label{tab:softmu_JETComp}
Number of events predicted by applying the soft muon misidentification matrix and observed in JET50 and JET100 data.}
\end{table}
}
{
Separate systematic uncertainties are estimated for the true muon identification efficiency and the misidentification rate. 
The invariant mass sideband subtraction technique used to obtain the muon efficiency matrix introduces uncertainties arising from the statistics of the 
$J/\psi$ sample. These uncertainties vary from 2\% - 70\%, depending on the bin in $p_T$ and $\eta$.
In addition, the maximum variation in efficiency of $8\%$ arising from the difference between isolated and non-isolated candidates is used
as an uncertainty representing the maximum possible difference between the $J/\psi$ sample environment and the $W/Z$ environment. This is added
in quadrature to the statistical uncertainty arising from the sideband subtraction method to obtain a final muon efficiency uncertainty of 8\% - 70\%.

The misidentification systematic uncertainty is obtained by selecting muon-free regions in samples triggered on high-$p_T$ jets and taking the difference between
observed and predicted soft muon misidentification rates. In this jet sample, at least 3 jets are required with $E_T>15$ GeV and $|\eta|<2.0$.  In order to reduce the contamination from real muons, any jet that contains an identified heavy quark decay is rejected, as is any track that has impact parameter significance $d_{0}/\sigma(d_{0})>2$.
In a sample having an online trigger requiring the presence of a jet with $E_T>100$ GeV, a difference of 4.8\% is observed between the observed and predicted soft muon identification rates.  A conservative estimate of twice this difference is used as the systematic uncertainty on the soft muon misidentification rate.

}

\ifthenelse{\boolean{Thesis}}
{\clearpage}
{}

\subsection{Application of soft lepton identification to $W/Z$ samples}
Additional selection criteria are applied to soft lepton candidates in the high $p_T$ $W$ and $Z$ boson data samples to reduce the amount of background 
in the search sample. Any track that is already identified as a high-$p_T$ electron or muon in the $W$ or $Z$ boson selection is ineligible to be identified as a soft muon. To reject badly measured tracks, each track is required to have at least one hit in the silicon detector.  For electron candidates, this hit is required to be within the first two layers of the silicon detector to help reject photon conversions. Each track is required to be inside of a reconstructed jet having $|\eta|<2.0$ and transverse energy of $E_{T}>5$ GeV, so that the heavy flavor fraction fit described later in Section~\ref{sec:frac_heavy} can be applied. (Note that the `jet' could be composed entirely of leptons, or even entirely of a single lepton.) Any track that is identified as a conversion partner is rejected. The track candidate must have a distance along the beamline $|\Delta z|<5$ cm from the high-$p_T$ trigger lepton. If the trigger lepton is the same flavor as the soft lepton, the invariant mass $M$ is calculated of the candidate and trigger, and the following mass ranges are rejected:
\begin{itemize}\setlength{\itemsep}{0pt}
\item $M<5$ GeV to suppress the $J/\psi$ and $b\bar{b}$ backgrounds.
\item $9<M<10$ GeV if the candidate track has opposite charge to the trigger lepton. This rejects $\Upsilon$ events.
\item $80<M<100$ GeV if the candidate track has opposite charge to the trigger lepton. This rejects $Z$ events.
\end{itemize} 

These additional selection criteria have a small effect on the benchmark model chosen for this analysis, cutting out 4.5\% of the signal leptons generated.

\section{Background prediction}
\label{sec:bg}

\ifthenelse{\boolean{Thesis}}
{The main SM backgrounds in the signal region of this analysis are semileptonic heavy quark decays and photons converting to electron-positron pairs.  The contributions from these backgrounds are estimated as described here.
}
{
The main SM backgrounds in this analysis are from $W$ + jets, Drell-Yan, QCD multijet, top quark, and diboson production processes.  The cross section and differential distributions of electroweak backgrounds from hard scattering processes are modeled using the {\sc Alpgen}~\cite{Alpgen} MC program, except for the top production and diboson production backgrounds, which are modeled by {\sc Pythia}~\cite{Pythia}.  {\sc Pythia} is used to model the parton showering in all samples.  These MC events are analyzed using a GEANT based detector simulation~\cite{CdfSim}. \ifthenelse{\boolean{Thesis}}{All of the MC datasets include the relevant K-factors in their normalization.}{}  The samples generated by {\sc Alpgen} are $W/Z + N_p$ partons (light flavor) and $W+ q \bar{q}+N_p$ partons, where $q = c, b$ (heavy flavor). The interface with the parton showering generates a double counting of heavy flavor events, which is corrected using the MLM matching method~\cite{MLM}.


The relative contributions from the various background sources can be seen qualitatively in Figure~\ref{fig:WZ_valid}. The cross sections used for every sample are described in~\cite{MyThesis}.  The final background predictions are summarized later in Section~\ref{sec:results}.  The QCD multijet background requires a different treatment since it is not possible to simulate it using MC. It is derived using data as explained below.

\subsection{QCD multijet background fraction}
\label{sec:Non_W}

The $W$ boson is identified by the presence of a high energy lepton and missing transverse
energy.  Events containing jets may emulate this signature; a dijet event, for example, may have
large $\met$ arising from the energy mismeasurement of one jet while the other jet in the event
can mimic an electron by leaving a track in the COT associated with an electromagnetic energy
deposit.  The contribution from these QCD multijet processes is estimated by using a data-derived model~\cite{Wbackground}.  This is accomplished by defining an object that is
similar to an electron, but has a much larger rate of contamination from jets; we refer to this as an ``anti-selected electron''.  An anti-selected electron is required to pass the same kinematic requirements as an electron, but must fail at least two of the identification requirements.

The number of events arising from the QCD multijet background is obtained by fitting the $\met$ distribution of the data
using two templates: an electroweak template obtained from $W+$ jets, $Z+$ jets and diboson MC, and a QCD template. 
The QCD template is obtained from the anti-selected electron sample after subtracting the expected $W$ boson contamination using the MC.  
The total number of events is kept constant and the fraction from each template is obtained from the fit.

After the fit is performed across the $\met$ distribution, the number of QCD events in the $W$ boson signal region is calculated by applying the selection of $\met>25$ GeV.  The MC electroweak contribution and the
data-derived QCD template are scaled to the result obtained from this $\met$ fit.  Figure~\ref{fig:MetFit} shows the result of this fit in the electron-triggered dataset.  A similar fit is performed in each muon-triggered dataset.  A systematic uncertainty of 26\% is applied to the QCD normalization, as found in~\cite{Wbackground}.

\begin{figure}[htb]
\begin{center}
\includegraphics[width=0.5\textwidth]{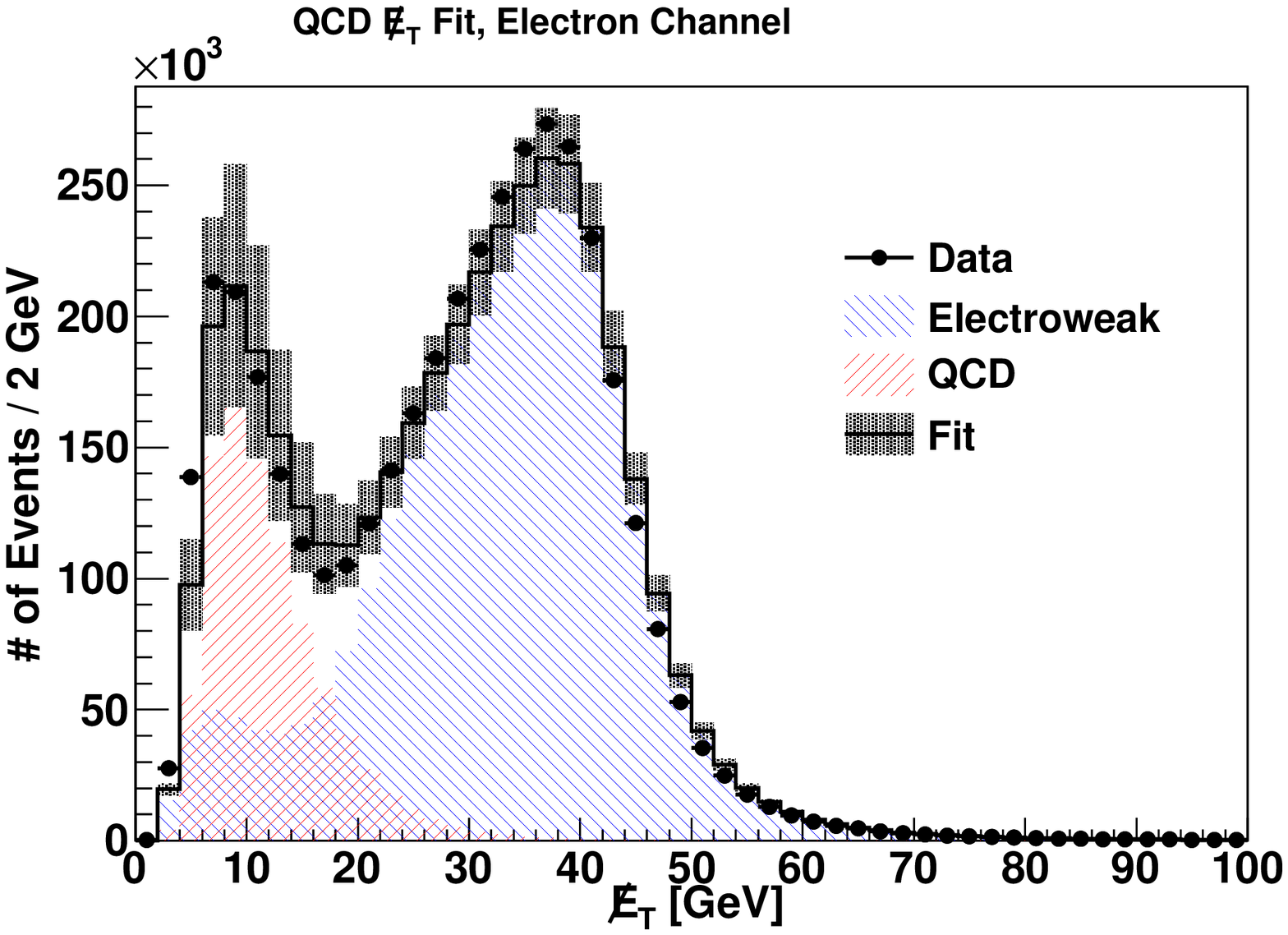}
\caption{
  \label{fig:MetFit}
  The fit to the $\met$ distribution of events with $m_{T}>20$ GeV and
$\Delta\phi(\met,l)>0.5$, in the electron-triggered dataset.  The ``electroweak'' template is obtained from Monte Carlo and the ``QCD'' 
template is obtained from the anti-selected electron data sample.  The systematic uncertainty of 26\% found in~\cite{Wbackground} is shown.
}
\end{center}
\end{figure}


} 
{}

\subsection{Heavy flavor background fraction}
\label{sec:frac_heavy}

\begin{figure*}
\begin{center}
%
\includegraphics[width=0.9\textwidth]{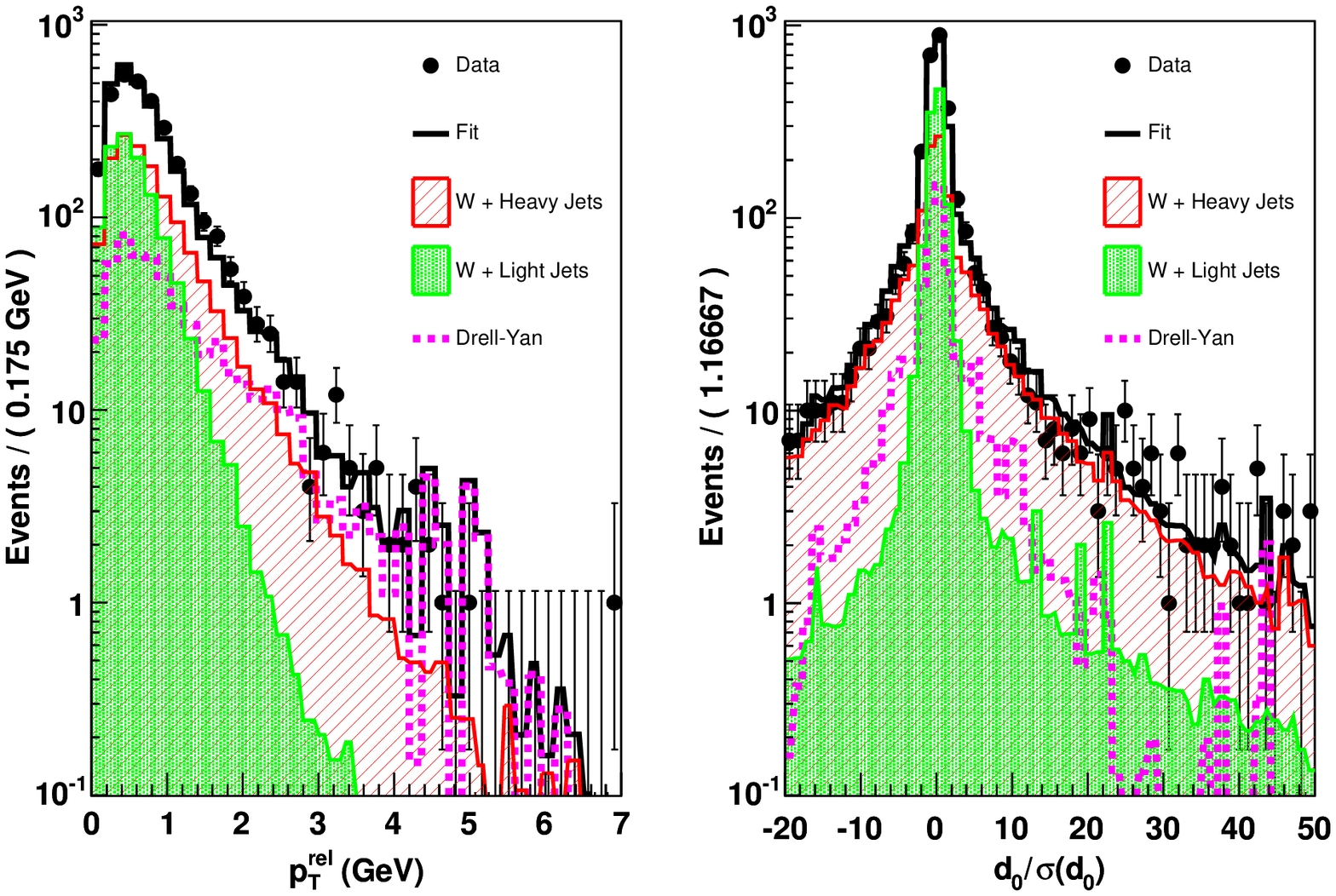}

\caption{\label{fig:pTrelFit}
The result of the simultaneous fit of the $W+1$ soft muon sample in the $p_{T}^{rel}$ and $d_0$ significance of the soft muon. The data distribution is fit to the sum of three components: $W+$heavy quark, $W+$light quark/gluon, and
Drell-Yan.
}
\end{center}
\end{figure*}

The leptonic decay of heavy flavor quarks creates a significant background contribution to the soft leptons of this analysis.  This background is estimated using the data in the $W$/$Z$ + exactly one soft muon channel, which should be dominated by SM processes.  A fit is performed in two distributions of soft muons which are sensitive to the heavy flavor fraction: $p_T^{rel}$, which is the momentum of the muon transverse to the direction of the jet in which it is found, and $d_0/\sigma(d_0)$, which is the significance of the muon's impact parameter with respect to the beamline.  A simultaneous fit is performed of these two distributions to a sum of templates from heavy flavor, light flavor, and Drell-Yan processes, as shown in Figure~\ref{fig:pTrelFit}.  These templates were acquired from the MC background samples. \ifthenelse{\boolean{Thesis}}{Only the contributions from light and heavy jets are allowed to vary, since the Drell-Yan cross section is already known to be well estimated from the $Z$ boson sample validation.

}{}  The result of this fit\ifthenelse{\boolean{Thesis}}{, shown in Table~\ref{tab:hf_fit},}{} is used to normalize the contributions of the three types of processes in the higher-multiplicity sample.  \ifthenelse{\boolean{Thesis}}{The results differ from the factor of $1.45 \pm 0.17$ found by previous analyses\cite{Method2} due to the much lower energy requirement for the jets in this analysis.}{}  The uncertainty resulting from the fit, ranging from 5\% to 34\% in the various samples, is used as a systematic uncertainty on this normalization.

\ifthenelse{\boolean{Thesis}}
{
\begin{table}
\begin{center}
\begin{tabular}{|l|c|c|}
\hline
Selection & Component & Scale Factor \\
\hline
TCE $W$ & Heavy & $2.51 \pm 0.20$ \\
 & Light & $0.93 \pm 0.05$ \\
CMUP $W$ & Heavy & $3.20 \pm 0.26$ \\
 & Light & $0.87 \pm 0.10$ \\
CMX $W$ & Heavy & $3.38 \pm 0.33$ \\
 & Light & $0.70 \pm 0.13$ \\
TCE $Z$ & Heavy & $5.21 \pm 0.71$ \\
 & Light & $0.75 \pm 0.15$ \\
CMUP $Z$ & Heavy & $3.76 \pm 0.97$ \\
 & Light & $1.33 \pm 0.23$ \\
CMX $Z$ & Heavy & $4.07 \pm 1.19$ \\
 & Light & $1.0 \pm 0.27$ \\
 \hline
\end{tabular}
\end{center}
\caption{\label{tab:hf_fit} The result of the fit to correct the heavy flavor fraction.}
\end{table}
}
{}

\subsection{Normalization of soft electron multiplicities}
\label{sec:normalization}

The heavy flavor fit described in Section~\ref{sec:frac_heavy} normalizes all of the data to the $W$/$Z$+$1 \mu$ channel.  However, we find a mismatch in the $W$/$Z$+$1 e$ channel, which has a large contribution from photon conversions. 
The difference between the predicted and observed numbers in the $W$/$Z$ plus exactly one electron channel is 34\% in the $W$ boson sample and 31\% in the $Z$ boson sample.  This is used as a systematic uncertainty for the normalization of all other MC with at least one additional identified electron~\cite{MyThesis}.

\section{Results}
\label{sec:results}

\ifthenelse{\boolean{Thesis}}
{
We perform a broad search for additional electrons and muons in the previously identified $W$ and $Z$ boson events. This signature of multiple leptons is common in many models of new physics with light mass scales and couplings to the electroweak sector.

First, a sample of 4,722,370 $W$ boson events and 342,291 $Z$ boson events is obtained from 5.1 fb$^{-1}$ of data.  In these base samples, good agreement with predictions is observed in all kinematic distributions, as shown in \Section{sec:validation}.

Then, techniques are demonstrated for soft electron and muon identification with no requirement of isolation.  For electrons, an efficiency of 80\% at $p_T=2$ GeV rising to 90\%-100\% for $4<p_T<20$ GeV is shown with a corresponding $2\%-4\%$ misidentification rate. For muons, an efficiency between 80\% at $p_T=3$ GeV and 40\% at $p_T=20$ GeV is shown with a misidentification rate less than 1\%.  These efficiencies are shown in Figures~\ref{fig:SE_compare} and \ref{fig:EffComp_softMuTag}.

\subsection{Soft lepton multiplicity}
}
{}

\begin{figure*}
\begin{center}
\ifthenelse{\boolean{Thesis}}{
\includegraphics[width=0.49\textwidth]{plots/slices/W_0.eps}
\hfil
\includegraphics[width=0.49\textwidth]{plots/slices/W_1.eps}

\includegraphics[width=0.49\textwidth]{plots/slices/W_2.eps}
\hfil
\includegraphics[width=0.49\textwidth]{plots/slices/W_3.eps}

\includegraphics[width=0.49\textwidth]{plots/slices/W_4.eps}
\hfil
\includegraphics[width=0.49\textwidth]{plots/slices/W_5.eps}
}
{
\includegraphics[width=0.45\textwidth]{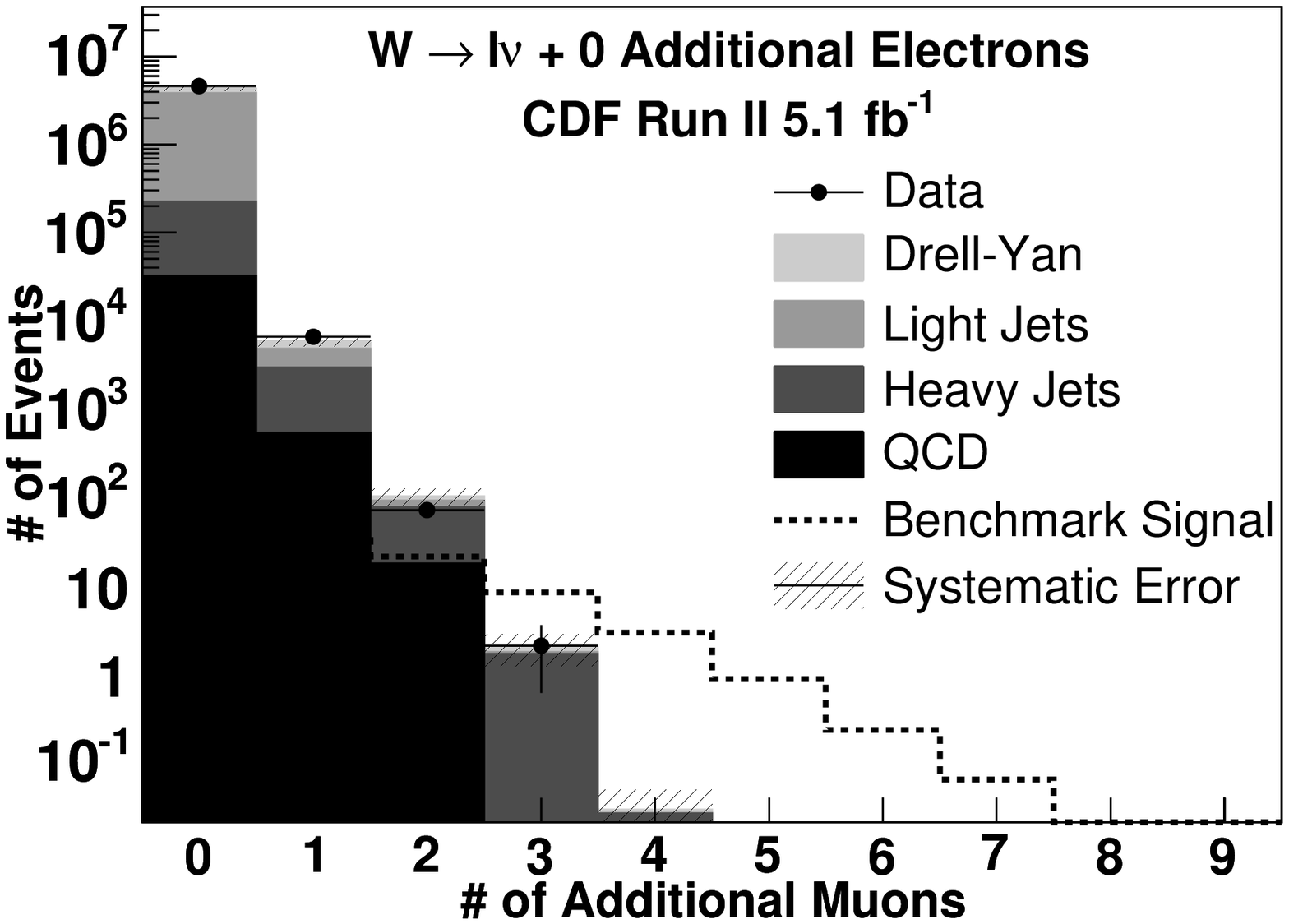}
\hfil
\includegraphics[width=0.45\textwidth]{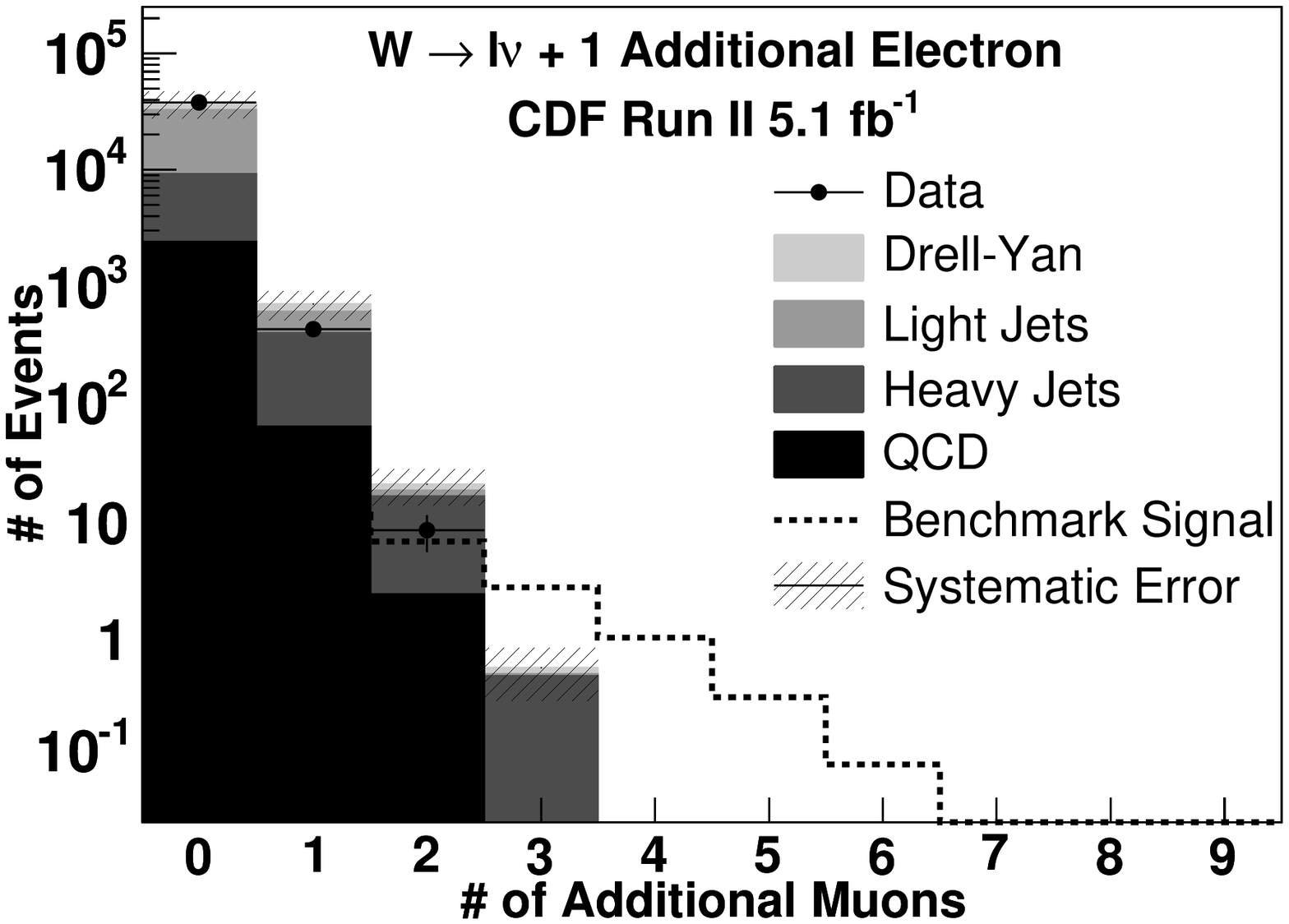}

\includegraphics[width=0.45\textwidth]{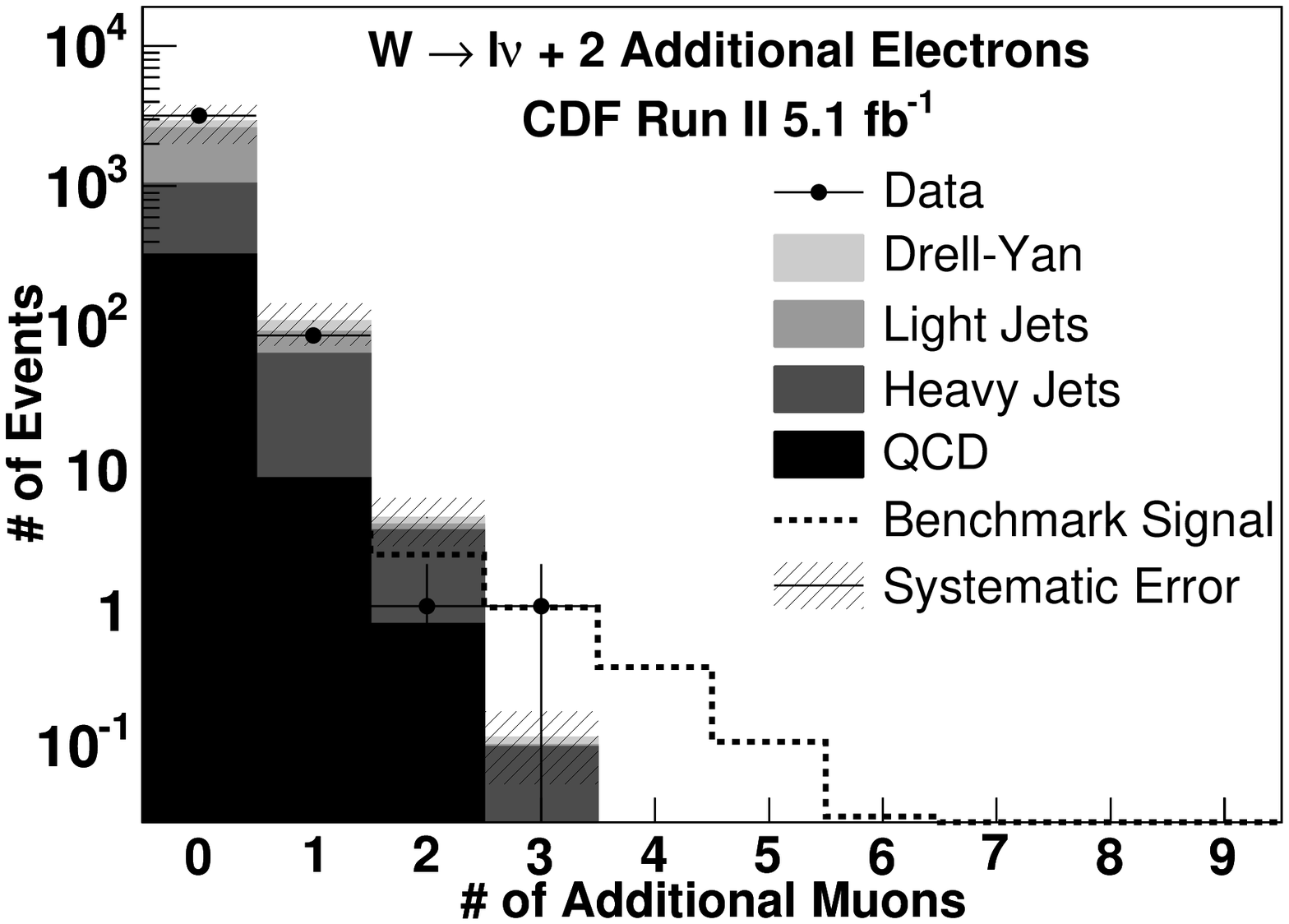}
\hfil
\includegraphics[width=0.45\textwidth]{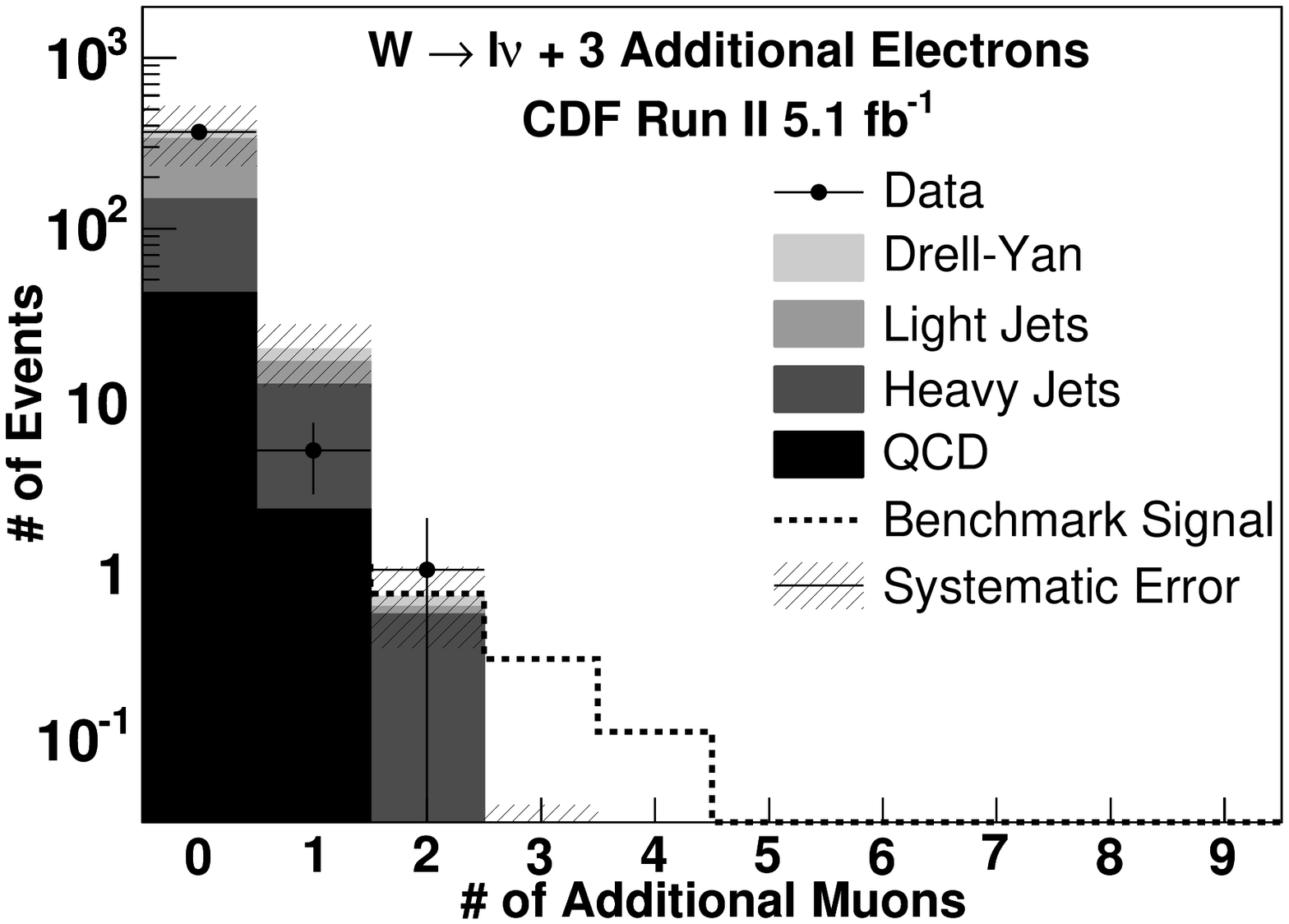}

\includegraphics[width=0.45\textwidth]{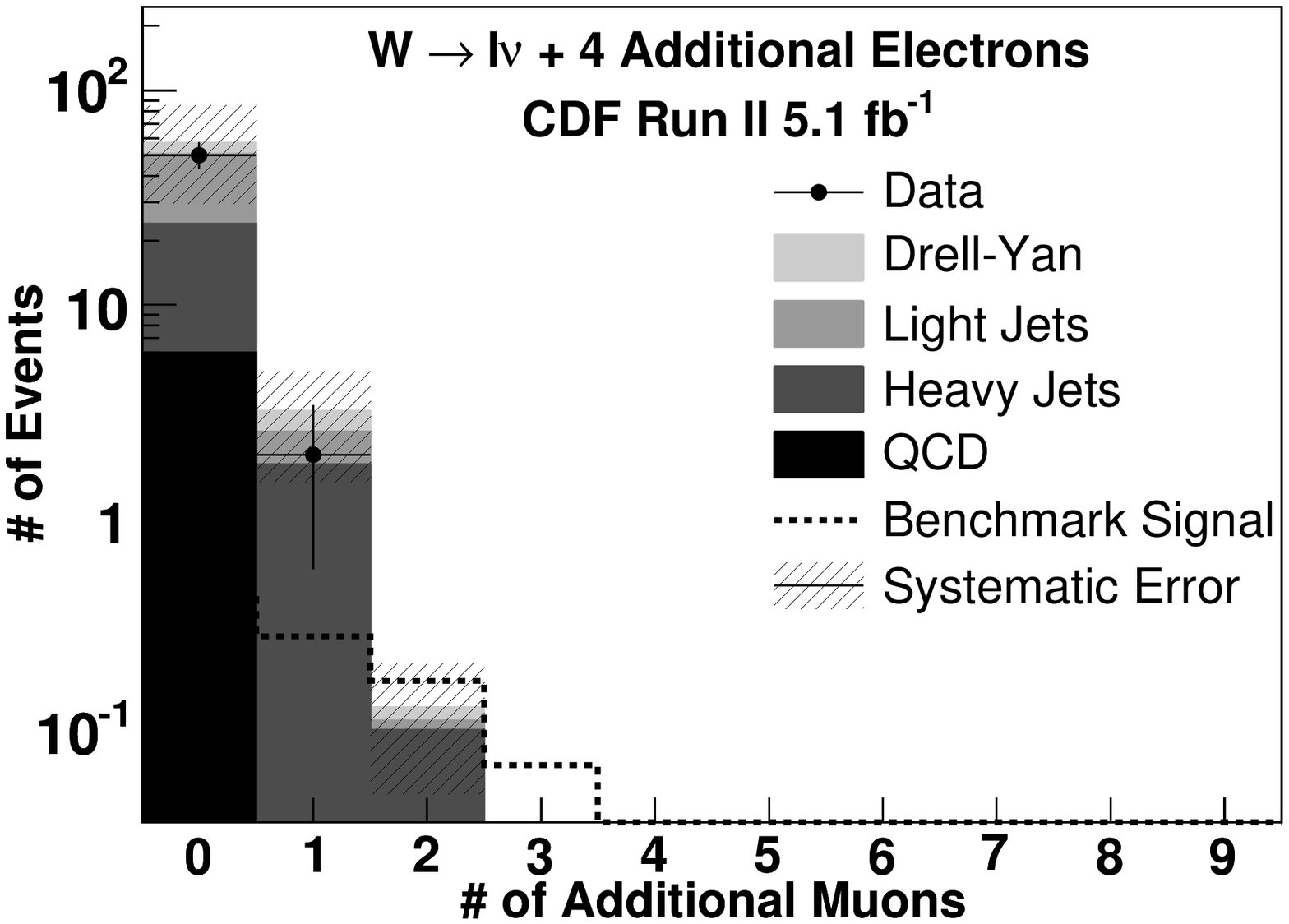}
\hfil
\includegraphics[width=0.45\textwidth]{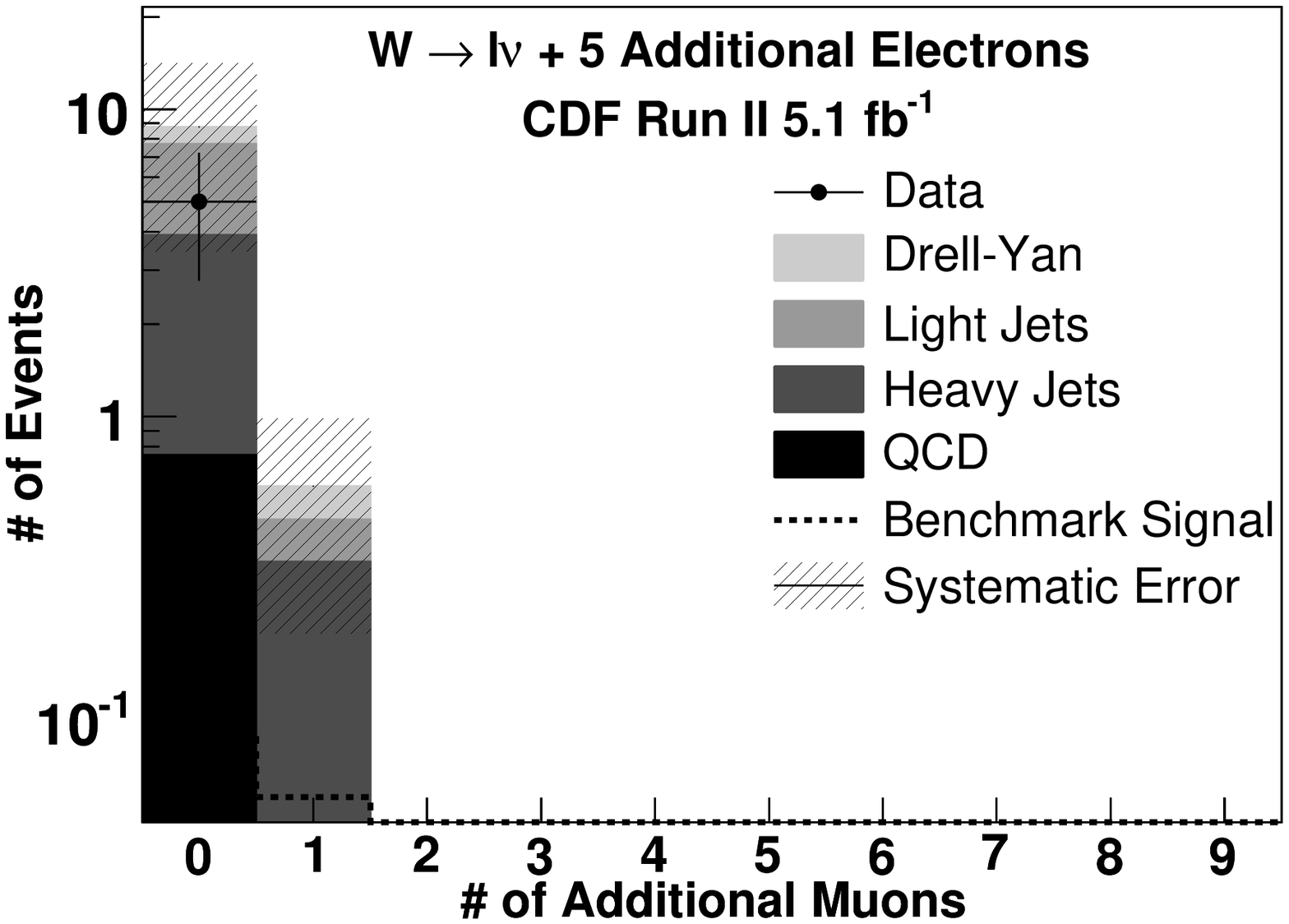}
}
\caption{\label{fig:Nem_W}
Multiplicity of additional electrons and muons after the $W$ boson selection.  The two-dimensional histogram of $N_\mu$ vs. $N_e$ is presented in slices of $N_e$ for ease of viewing.  Both hard
and soft leptons (but not the initial lepton used for the $W$ boson selection) are counted.  Note that the distributions combine
the electron- and muon-triggered events.}
\end{center}
\end{figure*}

\begin{figure*}
\begin{center}
\ifthenelse{\boolean{Thesis}}{
\includegraphics[width=0.49\textwidth]{plots/slices/Z_0.eps}
\hfil
\includegraphics[width=0.49\textwidth]{plots/slices/Z_1.eps}

\includegraphics[width=0.49\textwidth]{plots/slices/Z_2.eps}
\hfil
\includegraphics[width=0.49\textwidth]{plots/slices/Z_3.eps}

\includegraphics[width=0.49\textwidth]{plots/slices/Z_4.eps}
}
{
\includegraphics[width=0.45\textwidth]{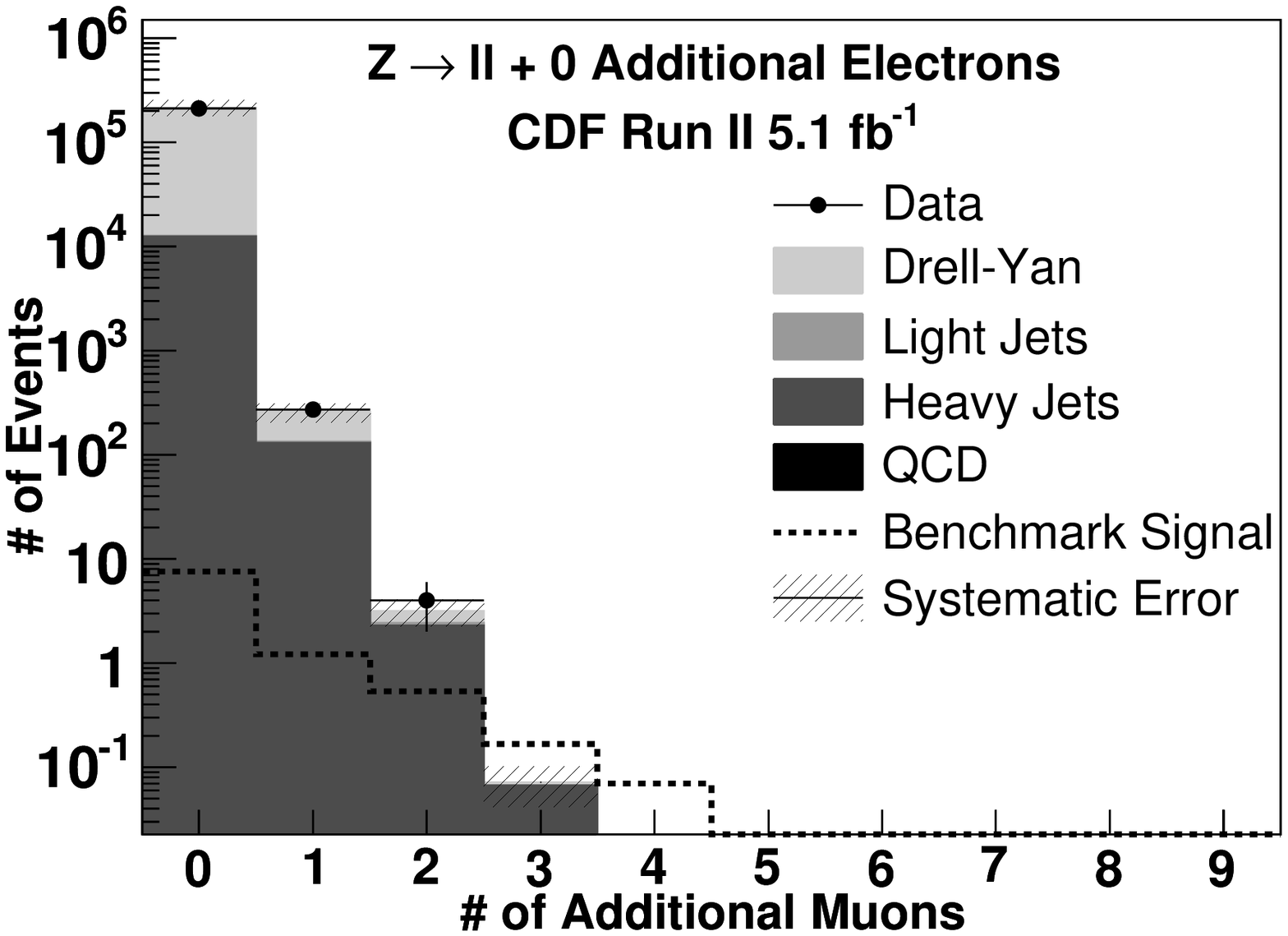}
\hfil
\includegraphics[width=0.45\textwidth]{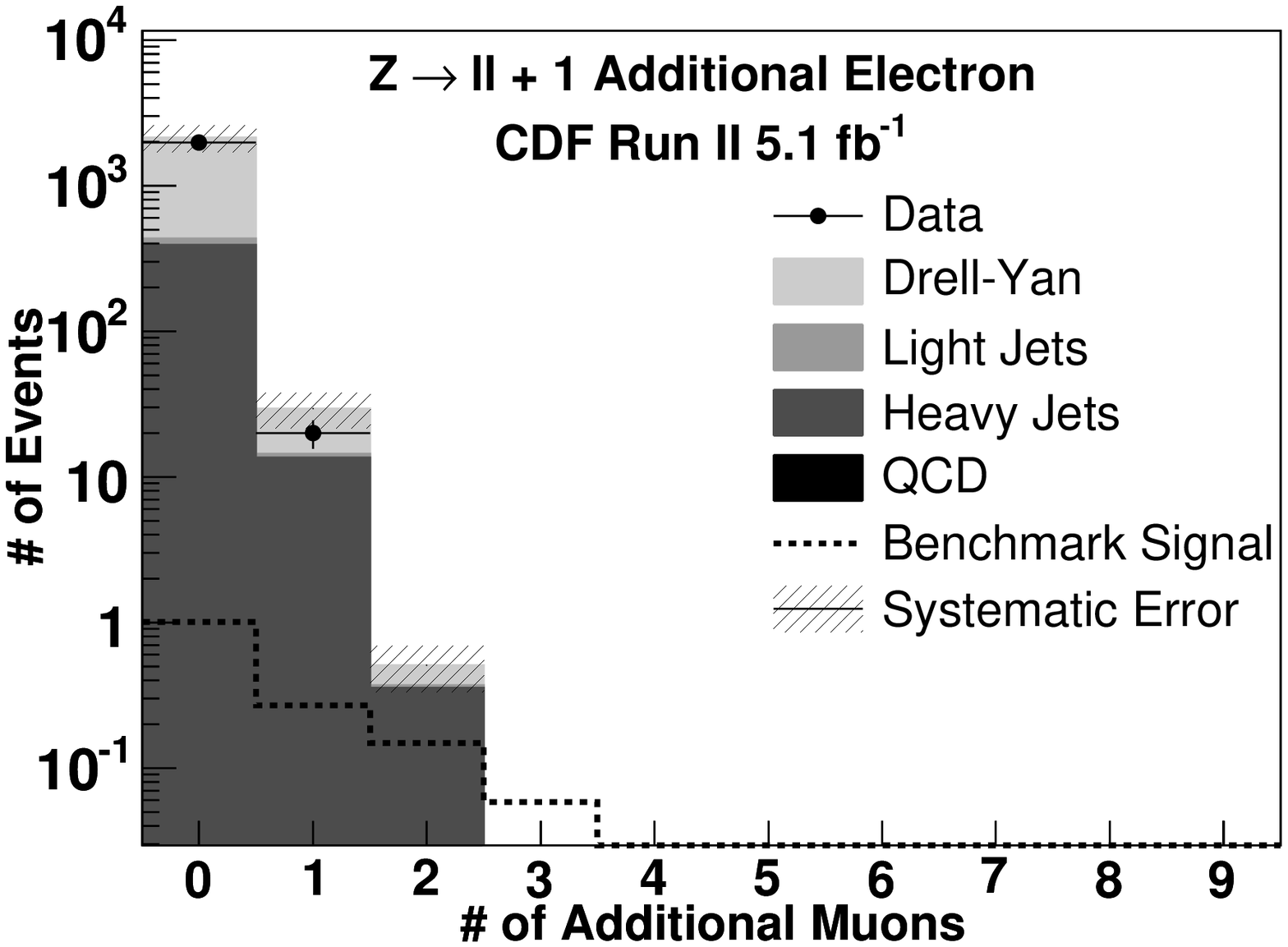}

\includegraphics[width=0.45\textwidth]{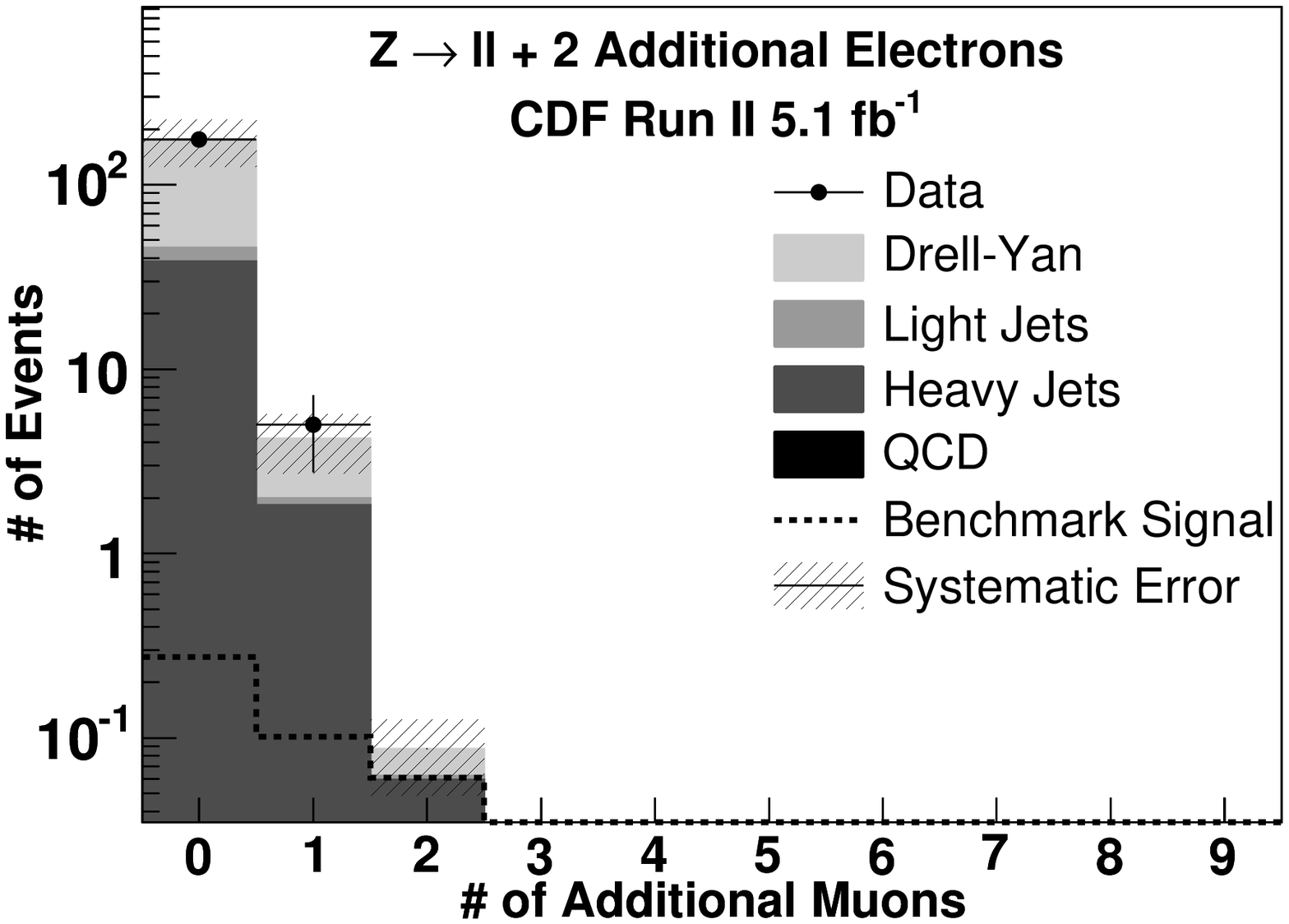}
\hfil
\includegraphics[width=0.45\textwidth]{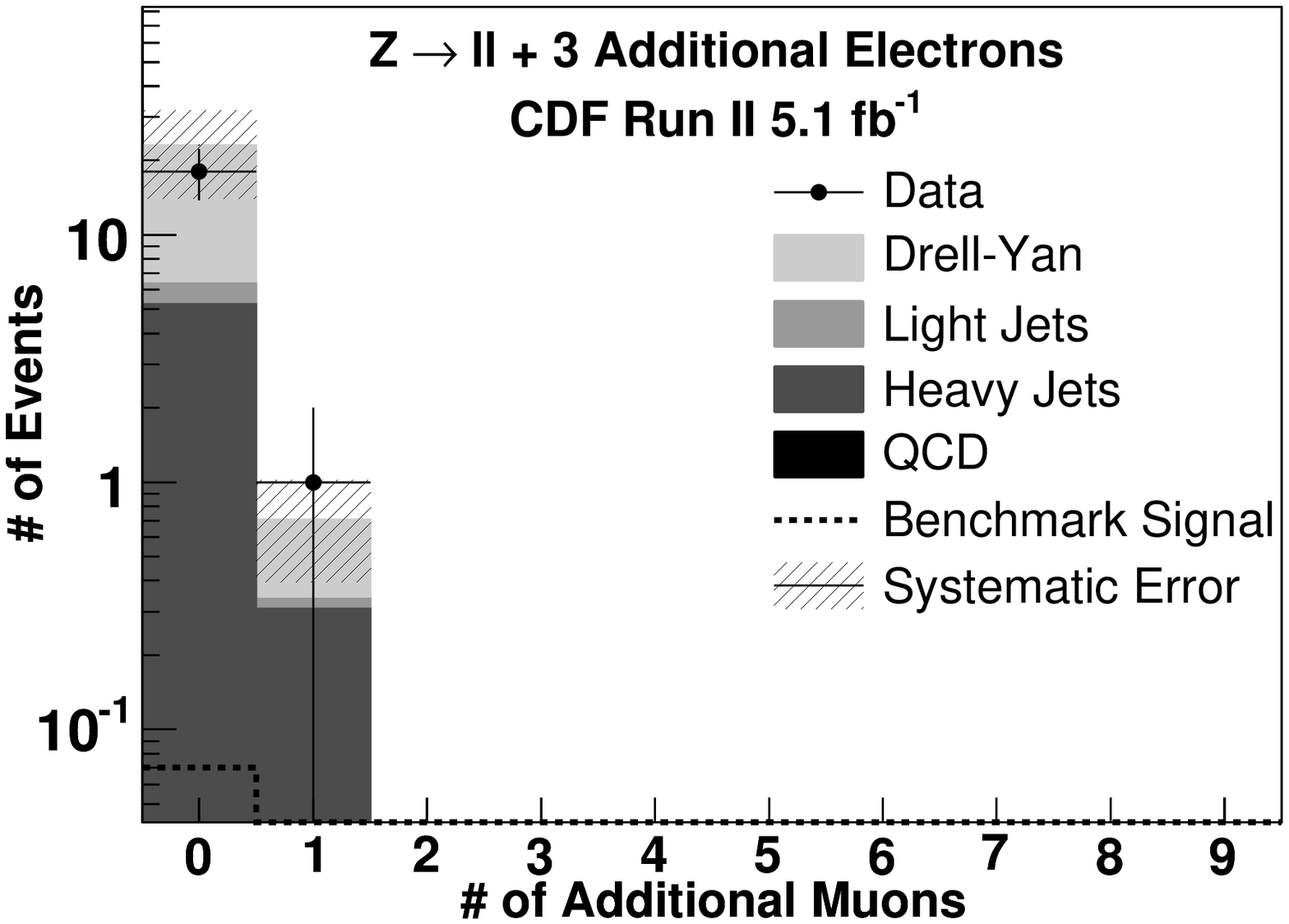}

\includegraphics[width=0.45\textwidth]{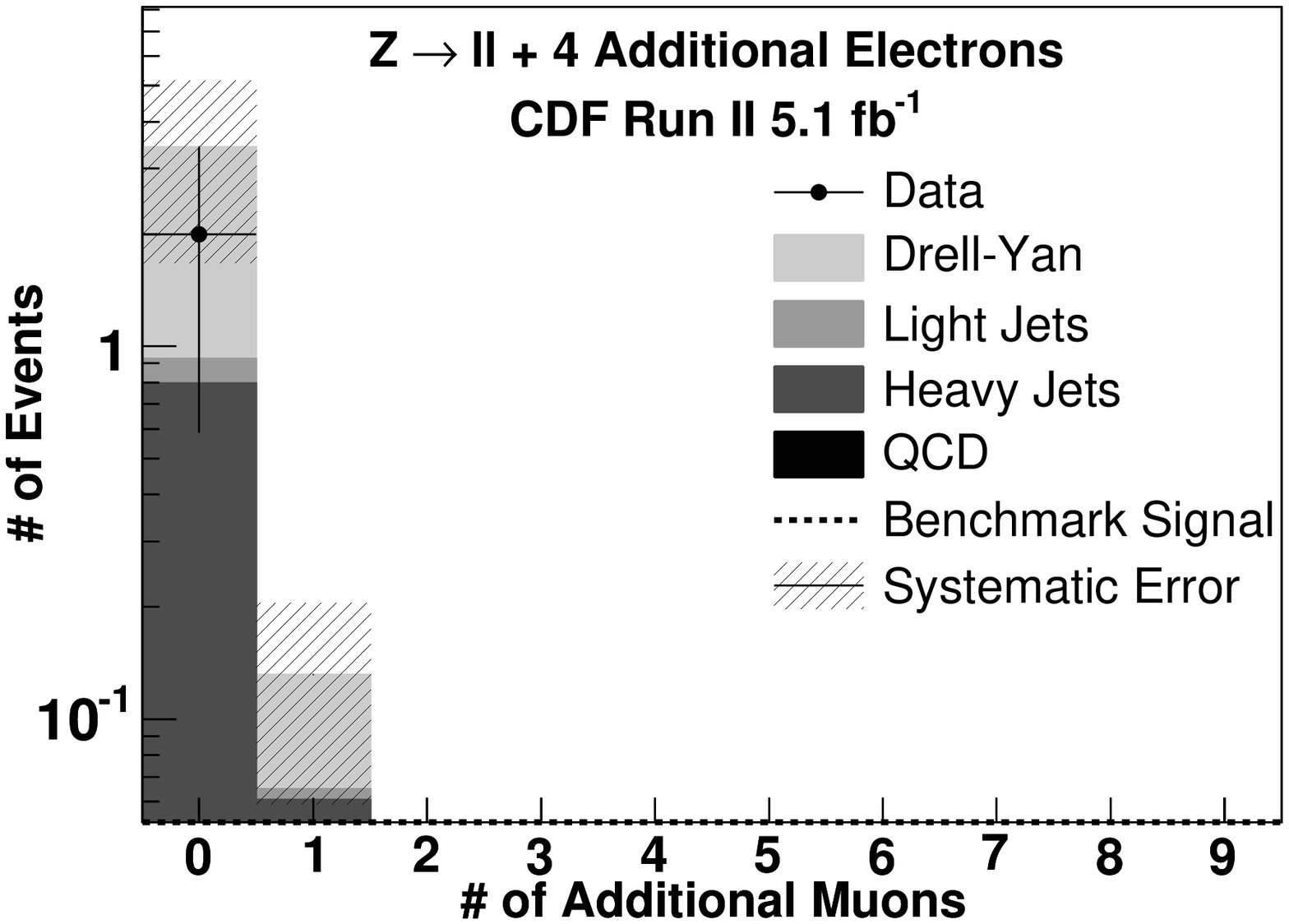}
}
\caption{\label{fig:Nem_Z}
Multiplicity of additional electrons and muons after the $Z$ selection.  The two-dimensional histogram of $N_\mu$ vs. $N_e$ is presented in slices of $N_e$ for ease of viewing.  Both hard
and soft leptons (but not the initial leptons used for the $Z$ boson selection) are counted.  Note that the distributions combine
the electron- and muon-triggered events.}
\end{center}
\end{figure*}

Using the soft lepton identification techniques described in Section~\ref{sec:SoftLep}, we count the numbers of $W$ and $Z$ boson events with multiple additional leptons. Figures~\ref{fig:Nem_W} and \ref{fig:Nem_Z} show the multiplicity of additional electrons ($N_e$) and muons ($N_\mu$) in these events, with the SM expectation and observed data overlaid.  The two-dimensional histograms of $N_\mu$ vs. $N_e$ are presented in slices of $N_e$ for ease of viewing.  These expected and observed event counts are also presented in Tables~\ref{tab:NResult_W} 
and~\ref{tab:NResult_Z} for ease of comparison with predictions from other models. The sources of systematic uncertainties are summarized in Table~\ref{tab:systematics}, with references to the sections in which they are described and evaluated.  Good agreement with the SM expectation is observed across the distributions.

\begin{table*}
\caption{\label{tab:NResult_W}
Summary of predicted and observed event counts by number of additional electrons ($N_e$) 
and muons ($N_{\mu}$) after the $W$ boson selection. The prediction of a model described in Section~\ref{sec:benchmark} is also shown for
comparison. Bins with less than 0.25 expected events in both signal and background and 0 observed events are not shown.}
\begin{center}
\begin{tabular}{ccccc}
\hline
\hline
$N_e$ & $N_\mu$ & Predicted SM Background & Predicted Dark Higgs Signal & Observed \\
\hline
0 & 0 & $4623512 \pm 315244$ & 158 & 4673896 \\
0 & 1 & $6463 \pm 807$ & 42 & 6498 \\
0 & 2 & $109 \pm 24$ & 21 & 70 \\
0 & 3 & $2.1 \pm 0.79$ & 8.0 & 2 \\
0 & 4 & $0.029 \pm 0.019$ & 2.8 & 0 \\
0 & 5 & $0.00026 \pm 0.00023$ & 0.83 & 0 \\
\\
1 & 0 & $46055 \pm 11387$ & 27 & 37778 \\
1 & 1 & $824 \pm 230$ & 11 & 425 \\
1 & 2 & $23 \pm 7.8$ & 6.4 & 8 \\
1 & 3 & $0.58 \pm 0.27$ & 2.6 & 0 \\
1 & 4 & $0.010 \pm 0.0074$ & 0.95 & 0 \\
1 & 5 & $0.00011 \pm 0.00011$ & 0.29 & 0 \\
\\
2 & 0 & $3600 \pm 1085$ & 7.1 & 3184 \\
2 & 1 & $129 \pm 43$ & 3.8 & 86 \\
2 & 2 & $4.9 \pm 1.8$ & 2.3 & 1 \\
2 & 3 & $0.13 \pm 0.067$ & 0.97 & 1 \\
2 & 4 & $0.0031 \pm 0.0024$ & 0.37 & 0 \\
\\
3 & 0 & $491 \pm 185$ & 1.9 & 366 \\
3 & 1 & $23 \pm 9.3$ & 1.2 & 5 \\
3 & 2 & $0.85 \pm 0.42$ & 0.72 & 1 \\
3 & 3 & $0.028 \pm 0.017$ & 0.30 & 0 \\
\\
4 & 0 & $79 \pm 38$ & 0.47 & 50 \\
4 & 1 & $3.9 \pm 2.1$ & 0.28 & 2 \\
\\
5 & 0 & $13 \pm 7.6$ & 0.096 & 5 \\
5 & 1 & $0.74 \pm 0.49$ & 0.058 & 0 \\
\\
6 & 0 & $2.0 \pm 1.5$ & 0.015 & 0 \\
\hline
\hline
\end{tabular}
\end{center}
\end{table*}

\begin{table*}
\caption{\label{tab:NResult_Z}
Summary of predicted and observed event counts by number of additional electrons ($N_e$) 
and muons ($N_{\mu}$) after the $Z$ selection. The prediction of a model described in Section~\ref{sec:benchmark} is also shown for
comparison. Bins with less than 0.25 expected events in both signal and background and 0 observed events are not shown.
}
\begin{center}
\begin{tabular}{ccccc}
\hline
\hline
$N_e$ & $N_\mu$ & Predicted SM Background & Predicted Dark Higgs Signal & Observed \\
\hline
0 & 0 & $215219 \pm 36886$ & 7.6 & 211448 \\
0 & 1 & $255 \pm 52$ & 1.2 & 270 \\
0 & 2 & $3.2 \pm 0.89$ & 0.54 & 4 \\
\\
1 & 0 & $2145 \pm 447$ & 1.0 & 1975 \\
1 & 1 & $30 \pm 8.1$ & 0.27 & 20 \\
1 & 2 & $0.51 \pm 0.18$ & 0.15 & 0 \\
\\
2 & 0 & $175 \pm 50$ & 0.28 & 176 \\
2 & 1 & $4.2 \pm 1.5$ & 0.10 & 5 \\
\\
3 & 0 & $23 \pm 9.0$ & 0.070 & 18 \\
3 & 1 & $0.71 \pm 0.31$ & 0.031 & 1 \\
\\
4 & 0 & $3.4 \pm 1.8$ & 0.019 & 2 \\
\\
5 & 0 & $0.52 \pm 0.35$ & 0.0044 & 0 \\
\hline
\hline
\end{tabular}
\end{center}
\end{table*}

\begin{table*}
\caption{\label{tab:systematics} Sources of systematic uncertainties.  Their size is measured both as a percentage and as the number of events in a benchmark-signal-rich region, defined as a $W$ or $Z$ boson plus at least 3 additional muons with $p_T>3$ GeV.  Note that, although some of the systematics are large, they have little effect in the signal region due to there being negligible SM background. 
}
\begin{center}
\begin{tabular}{lcc}
\hline
\hline
Systematic Source & Size & Effect in Large S/B Region (Events)\\
\hline
Trigger Efficiency \ifthenelse{\boolean{Thesis}}{(Sec.~\ref{sec:WZ_R})}{\cite{MyThesis}} & $\pm (1.6$ - $5.9)$\% & $\pm 0.06$ \\
QCD fraction (Sec.~\ref{sec:Non_W}) & $\pm26$\% & 0 \\
Soft $e$ real rate (Sec.~\ref{sec:se_validation}) & $\pm15$\% & $\pm 0.04$ \\
Soft $e$ fake rate (Sec.~\ref{sec:se_validation}) & $\pm15$\% & $\pm 0.11$ \\
Soft $\mu$ real rate (Sec.~\ref{sec:sm_syst}) & $\pm$(8-70)\% & $\pm 0.64$ \\
Soft $\mu$ fake rate (Sec.~\ref{sec:sm_syst}) & $\pm10$\% & $\pm 0.34$ \\
Soft $e$ normalization (Sec.~\ref{sec:normalization}) & $\pm$(31-39)\% & $\pm 0.24$ \\
Heavy Flavor Fraction (Sec.~\ref{sec:frac_heavy}) & $\pm$(5-34)\% & $\pm 0.25$ \\
\hline
\hline
\end{tabular}
\end{center}
\end{table*}

In particular, very few multi-muon events are observed.  This is the region where many lepton jet models would be expected to show an excess, since a potential signal in the multi-electron region would be more likely to be hidden by the large background contribution from photon conversions.  Only three events containing 3 muons beyond the $W$ selection are observed, which is consistent with the SM expectation of 2.9 events.  No events are observed containing four or more additional muons.
\ifthenelse{\boolean{Thesis}}{A summary of the high lepton multiplicity events that are observed in the data is shown in Appendix~\ref{sec:appendix_multi_lepton}.}{}

\ifthenelse{\boolean{Thesis}}
{
\subsection{Soft lepton kinematics}
Many new physics scenarios, in addition to creating excesses in the soft lepton multiplicity, would result in discrepancies in the soft lepton kinematics.  For example, in the benchmark model described in Table~\ref{tab:benchmark_params}, pairs of leptons are produced by decaying dark photons, which would create a mass peak at 300 MeV in the dilepton mass.

With no excess having been seen in the soft lepton multiplicity, the dilepton mass distributions are shown in Figure~\ref{fig:SL_mass}.  All distribution are consistent with the SM predictions.  The most discrepant is the $Z+\mu\mu$ mass distribution, but the discrepancy does not pass the 95\% confidence level.

\begin{figure}
\includegraphics[width=0.49\textwidth]{plots/m_sl_sl/W_se_se.eps}
\hfil
\includegraphics[width=0.49\textwidth]{plots/m_sl_sl/Z_se_se.eps}

\includegraphics[width=0.49\textwidth]{plots/m_sl_sl/W_se_sm.eps}
\hfil
\includegraphics[width=0.49\textwidth]{plots/m_sl_sl/Z_se_sm.eps}

\includegraphics[width=0.49\textwidth]{plots/m_sl_sl/W_sm_sm.eps}
\hfil
\includegraphics[width=0.49\textwidth]{plots/m_sl_sl/Z_sm_sm.eps}
\caption{\label{fig:SL_mass} Distributions of the invariant mass of each pair of soft leptons $ee$ (top), $e\mu$ (center), and $\mu\mu$ (bottom).  The $W$-selected events are on the left and the $Z$-selected events are on the right. Note that the distributions combine the electron-and muon-triggered events.  The contribution from conversions swamps any new physics signal in the $m(e,e)$ and $m(e,\mu)$ distributions, but the $m(\mu,\mu)$ distribution (bottom) is sensitive to the benchmark model as well as to other new physics models.}
\end{figure}
}
{}

\subsection{Benchmark model}
\label{sec:benchmark}

This is a general signature-based search, and as such is applicable to many different models.  We choose an example model from the representative lepton jet models presented in Ref.~\cite{HiggsToLeptonJets}.  The benchmark model chosen for this analysis is an adaptation of the `Neutralino Benchmark Model,' in which the Higgs decays principally to a pair of the lightest supersymmetric particles, which then decay through a dark sector to lepton jets.  A MC sample of signal events was generated from this model using {\sc Pythia}.  The signal from this model to which this analysis is most sensitive is associated production of a $W$ or $Z$ boson and a Higgs boson, which has a cross section of 389 fb.  This cross section would result in 1647 $W+$ Higgs events and 322 $Z+$ Higgs events in the data sample of this analysis before applying any selection criteria.

The particular parameters of the model~\cite{ReecePrivate} were chosen to create a `typical' model of this class. The MSSM parameters ($\mu$, $m_1$, $m_2$, $\tan(\beta)$ and $\sin(\alpha)$) avoid previous limits from searches for supersymmetry while making the lightest supersymmetric partner ($\chi_0$) the favored Higgs decay channel.  The Higgs has a mass near that favored by precision measurements.  The branching fractions for $\chi_0$ decaying into the dark neutralinos ($\chi_d$) and dark photons ($\gamma_d$) simply model the sort of cascade decay illustrated in Figure~\ref{fig:HHFeynman}. The mass of the dark photon is chosen in order to make the additional leptons that are produced approximately half muons and half electrons.  These parameters are summarized in Table~\ref{tab:benchmark_params}.


\begin{table}
\caption{\label{tab:benchmark_params}
Parameters used for the benchmark model based on that in Ref.~\cite{HiggsToLeptonJets}.  The first five parameters are the inputs to the MSSM including the branching fractions
for $\chi_0 \to \chi_d + N \gamma_d$~\cite{ReecePrivate}.
}
\begin{center}
\begin{tabular}{cc}
\hline
\hline
Parameter & Value \\
\hline
$\mu$ & 149 GeV \\
$m_1$(bino) & 13 GeV \\
$m_2$(wino) & 286 GeV \\
$\tan(\beta)$ & 3.5 \\
$\sin(\alpha)$ & --0.28 \\
$m_{\chi_0}$ & 10 GeV \\
$m_H$ & 120 GeV \\
$m_{\chi_d}$ & 1 GeV \\
$m_{\gamma_d}$ & 300 MeV \\
BR$(\chi_0 \to \chi_d + 2 \gamma_d)$ & 33\% \\
BR$(\chi_0 \to \chi_d + 3 \gamma_d)$ & 33\% \\
BR$(\chi_0 \to \chi_d + 4 \gamma_d)$ & 33\% \\
\hline
\hline
\end{tabular}
\end{center}
\end{table}



We set a 95\% confidence level limit on the production of this benchmark model.  The limit is set at $0.312\times\sigma$, or 112 fb.  The model can be ruled out at the standard cross section at a confidence level of 99.7\%. 
Both of these limits are set in the Bayesian framework using the {\sc mclimit} tools~\cite{mclimit} running over the combined $W$ and $Z$ channels in Figures~\ref{fig:Nem_W} and \ref{fig:Nem_Z} (Tables \ref{tab:NResult_W} and \ref{tab:NResult_Z}).



\subsection{Application to other models}

In addition to the benchmark model discussed in Section~\ref{sec:benchmark}, limits can be set on a wide range of alternate models.  A rough estimate of the limit for a particular model can be made by normalizing its production to the $W$ or $Z$ boson cross section, applying the efficiencies in 
Tables~\ref{tab:se_eff} and \ref{tab:sm_eff}
to the additional leptons, and comparing the result to the observed and predicted numbers of additional leptons in Tables~\ref{tab:NResult_W} and \ref{tab:NResult_Z}.  For ease of reference, a summary of the kinematic selections for identified objects is presented in Table~\ref{tab:ObjectEfficiencies}.

\begin{table*}
\caption{\label{tab:ObjectEfficiencies} Summary of kinematic requirements to find various objects.  These numbers can be used to set limits on many models that predict production of additional leptons.}
\begin{center}
\begin{tabular}{ccc}
\hline
\hline
Object & Requirements & Number Observed \\
\hline
$W$ & $p_T(e/\mu) > 20$ GeV & 4,722,370 \\
 & $|\eta(e)| < 1.1$, $|\eta(\mu)| < 1.5$ & \\
 & $\met > 25$ GeV & \\
 & $m_T(l,\met) > 20$ GeV & \\
 & $d\phi(l, \met) > 0.5$ & \\
\\
$Z$ & $p_T(e/\mu) > 20$ GeV & 342,291 \\
 & $p_T(e_2) > 12$ GeV, $p_T(\mu_2) > 10$ GeV & \\
 & $|\eta(e)| < 1.1$, $|\eta(\mu)| < 1.5$ & \\
 & 76 GeV $< m(l_1, l_2) <$ 106 GeV & \\
\\
soft $e$ & $p_T(e) > 2$ GeV & See Tables~\ref{tab:NResult_W} and \ref{tab:NResult_Z} \\
 & $|\eta(e)| < 1$ & \\
 & $\mathcal{L} > 0.99$ (Efficiency in Table~\ref{tab:se_eff}) & \\
\\
soft $\mu$ & $p_T(\mu) > 3$ GeV & See Tables~\ref{tab:NResult_W} and \ref{tab:NResult_Z} \\
 & $|\eta(\mu)| < 1.5$ & \\
 & $|\mathcal{L}| < 3.5$ (Efficiency in Table~\ifthenelse{\boolean{Thesis}}{\ref{fig:EffComp_softMuTag}}{\ref{tab:sm_eff}}) & \\
\hline
\hline
\end{tabular}
\end{center}
\end{table*}

In general, any model that predicts significant numbers of 3-muon events can be ruled out, since only three such events are observed in the sample, consistent with the SM background.  However, models that produce multiple electrons can more easily be accommodated, since photon conversions result in a much higher background in that region.

\section{Conclusions}
\label{sec:conclusions}

This analysis expands the reach of previous searches for additional leptons by allowing leptons to be reconstructed from a much lower $p_T$ threshold and with no requirement of isolation.  This greatly increases the acceptance to find lepton jets or similar excesses of leptons from effects beyond the SM.  No indication of such new effects is seen in the data sample.  A 95\% confidence level limit is set on an example benchmark model of supersymmetric Higgs production, and a framework is provided to set limits on a class of other models.

\begin{acknowledgments}
We thank Lian-Tao Wang and Matthew Reece for assistance in implementing the dark sector Higgs model and guidance in how to best formulate limits for this model.

We also thank the Fermilab staff and the technical staffs of the participating institutions for their vital contributions. This work was supported by the U.S. Department of Energy and National Science Foundation; the Italian Istituto Nazionale di Fisica Nucleare; the Ministry of Education, Culture, Sports, Science and Technology of Japan; the Natural Sciences and Engineering Research Council of Canada; the National Science Council of the Republic of China; the Swiss National Science Foundation; the A.P. Sloan Foundation; the Bundesministerium f\"ur Bildung und Forschung, Germany; the Korean Science and Engineering Foundation and the Korean Research Foundation; the Science and Technology Facilities Council and the Royal Society, UK; the Institut National de Physique Nucleaire et Physique des Particules/CNRS; the Russian Foundation for Basic Research; the Ministerio de Ciencia e Innovaci\'{o}n, and Programa Consolider-Ingenio 2010, Spain; the Slovak R\&D Agency; and the Academy of Finland. 
\end{acknowledgments}

\ifthenelse{\boolean{Thesis}}
{\clearpage
\addcontentsline{toc}{chapter}{Bibliography}}
{}

\clearpage


\begin{thebibliography}{99}
\bibitem{EtPt}
We use the convention that ``momentum'' refers to $pc$ and ``mass'' to
$mc^2$.
\end{thebibliography}
\end{document}